# A dual-scale stochastic analysis framework for creep failure considering microstructural randomness

Weichen Kong [a], Yanwei Dai [b,*], Xiang Zhang [c], Yinghua Liu [a,*]

[a] Department of Engineering Mechanics, AML, Tsinghua University, Beijing 100084, China

[b] Institute of Electronics Packaging Technology and Reliability, Department of mechanics, Beijing University of Technology, Beijing 100124, China

[c] Department of Mechanical Engineering, University of Wyoming, Laramie, WY 82071, USA

[*] Corresponding authors:

E-mail addresses: ywdai@bjut.edu.cn (Yanwei Dai), yhliu@mail.tsinghua.edu.cn (Yinghua Liu).

# Abstract

Creep failure under high temperatures is a complex multiscale and multi-mechanism issue involving inherent microstructural randomness. To investigate the effect of microstructures on the uniaxial/multiaxial creep failure, a dual-scale stochastic analysis framework is established to introduce the grain boundary (GB) characteristics into the macroscopic analysis. The nickel-base superalloy Inconel 617 is considered in this study. Firstly, the damage mechanisms of GBs are investigated based on the crystal plasticity finite element (CPFE) method and cohesive zone model (CZM). Subsequently, based on the obtained GB damage evolution, a novel Monte Carlo (MC) approach is proposed to establish the relationship between the GB orientation and area distribution and macroscopic creep damage. Finally, a dual-scale stochastic multiaxial creep damage model is established to incorporate the influence of the random GB orientation and area distribution. With the numerical application of the proposed creep damage model, the random initiation and growth of creep cracks in the uniaxial tensile specimen and the pressurized tube are captured and analyzed. The proposed stochastic framework effectively considers the inherent randomness introduced by GB characteristics and efficiently realizes full-field multiscale calculations. It also shows its potential applications in safety evaluation and life prediction of creep components and structures under high temperatures.

# 1. Introduction

Creep is an essential factor to consider for materials and structures subjected to long-term high-temperature conditions (Drexler et al., 2018; Joy et al., 2024; Kumar and Capolungo, 2022; Xiao et al., 2021; Xu et al., 2021), which has received sustained attention over an extended period for different materials (Bartošák and Horváth, 2024; Ding et al., 2022; Li et al., 2023; Skamniotis et al., 2023; Xu et al., 2022b). The creep deformation and failure are controlled by different multiscale mechanisms, such as vacancy diffusion (Liang et al., 2024), dislocation glide and climb (Galindo-Nava et al., 2023; Kim et al., 2016b; Ling et al., 2023; Song et al., 2024; Xiao et al., 2019; Xiao et al., 2021), grain boundary (GB) sliding (Kim et al., 2016b), precipitation (Drexler et al., 2018; Salvini et al., 2024), creep voids (Bieberdorf et al., 2021; Kumar and Capolungo, 2022), etc. This complex creep deformation and failure mechanism presents challenges for structural failure analysis and safety assessment. Particularly for critical materials in next-generation clean energy systems, such as Inconel 617 (Choi et al., 2022; Kan et al., 2019; Kim et al., 2010; Tung et al., 2014), the multiaxial creep damage occurring during long-term service exhibits intricate micro-to-macro scale correlations.

Specifically, for long-term creep behavior under high temperature and low stress, failure caused by typical creep voids on GBs widely exists (Kim et al., 2015; Wang et al., 2021). The growth of such GB creep voids is often influenced by GB diffusion. The pioneering work was carried out by Cocks and Ashby (1982) for typical creep void, providing an approximate equation for the rate of cavity growth. Based on the cavity growth model, Wen et al. (2017) investigated the crack growth behavior under monotonic and cyclic loading. Similarly, Sanders et al. (2017) effectively analyzed the creep fracture of high-temperature alloys under uniaxial and biaxial stress conditions. Overall, these studies have yet to consider the impact of crystalline microstructure on creep failure.

To analyze the complex creep mechanisms from a microscopic perspective, multiscale modeling and computational strategies such as Crystal Plasticity Finite Element (CPFE), Cohesive Zone Model (CZM), and Phase Field Method (PFM) have been widely employed. Phan et al. (2017) investigated the uniaxial creep deformation and rupture mechanisms of Inconel 617 through CPFE-CZM framework. Zhang et al. (2020) analyzes Type IV failure in weldments of creep strength enhanced ferritic steel using a novel integrated microstructure- and micromechanics-based finite element model. Li et al. (2023) numerically studied the creep-fatigue damage mechanisms and estimated the crack initiation life through the CPFE method and GB damage model. Salvini et al. (2024) revealed the influence of intergranular carbide precipitates on creep failure through a proposed coupled CPFE-PFM framework. However, these deterministic mechanical models at the microscale are typically applied to Representative Volume Element (RVE) analyses or primarily focus on hotspot regions such as notch roots, crack tips and welded joints.

Moreover, incorporating microscopic creep mechanisms into the macroscopic deformation and failure analysis of materials and structures, particularly their full-field response, remains a challenging task. Taking Inconel 617 as a representative case, comprehensive experimental testing (Kim et al., 2016a; Kim et al., 2015; McMurtrey, 2017; Narayanan et al., 2017; Tung et al., 2014; Wright and wright, 2013; Wright et al., 2014; Wright, 2021) and microstructural characterization analyses (Bagui et al., 2022; Kan et al., 2019; Lillo and Wright, 2015; Sharma et al., 2009; Tung et al., 2014; Wang et al., 2022a; Wang et al., 2022b; Wang et al., 2021; Zhang and Oskay, 2016) have been conducted. These studies have addressed critical aspects including multiaxial creep failure analysis (Sanders et al., 2017; Tung et al., 2014), damage modeling (Choi et al., 2022), and lifetime prediction (Kan et al., 2019) for Inconel 617. Therefore, this material serves as an ideal benchmark for bridging microscopic mechanisms to macroscopic performance due to its well-documented failure modes and extensive experimental datasets in crucial high-temperature applications.

In practice, to realize a multiscale simulation on failure of materials, specific calculation frameworks are required (Feyel, 2003; Hernández et al., 2014). As shown in Fig. 1, the microstructures corresponding to the material points of the macroscopic structure (here assumed to be equivalent to the integration points of the elements) are different from each other. These microstructures are shown in Fig. 1b and simulated at the microscale based on the deterministic constitutive model $f$. Then, the micro information from micro simulation is transmitted to the macro structures (the red dashed arrow in Fig. 1). However, conducting detailed microscale simulations at the microscale is often constrained by high computational costs, typically limiting analyses to localized hotspot regions. To overcome the computational bottleneck, a stochastic modeling framework may enhance the efficiency and enable full-field assessments of entire components. As shown in Fig. 1c, the macroscopical responses of RVEs at material points are obtained through extensive microscopic calculations. It is assumed that the macroscopical deformation and failure properties conform to a certain statistical distribution. The statistical characteristics derived are then reflected in the macroscopic constitutive relation $F$ for macroscopic computations. It is noted that the constitutive relation $F$ is a symbol representing the material properties of both deformation and failure. In this framework, Due to the consideration of the microstructure, the material properties in the constitutive relation $F$ are stochastic variables rather than deterministic ones. The final calculations are conducted at the macroscopic scale. Specific microstructures are no longer considered; instead, their effects are represented and characterized solely through statistical characteristics of macroscopic constitutive parameters.

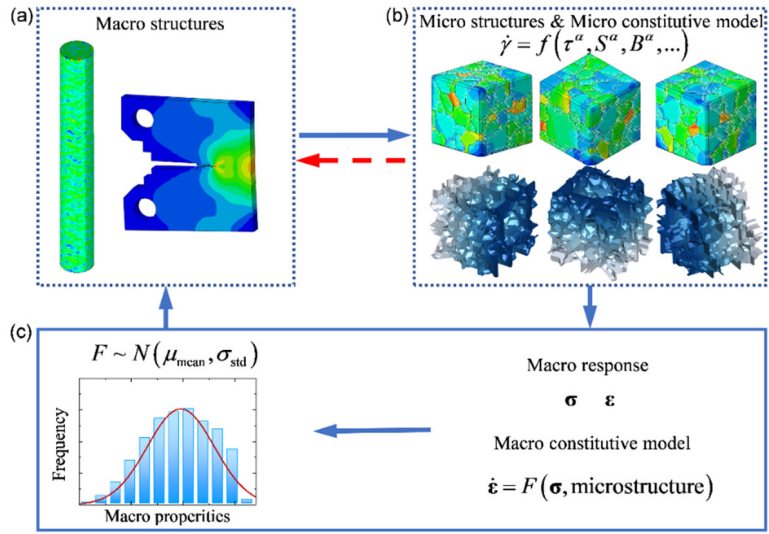

Fig. 1 Schematic illustration of the randomness in macroscopic behavior induced by the inherent randomness in microstructural features

Quantifying the stochastic characteristics of macroscopic constitutive parameters is of significant value for understanding the complex failure behaviors of materials and conducting reliable assessments (EDF Energy Nuclear Generation Ltd, 2018; Wang et al., 2023). On one hand, the stochastic characteristics of macroscopic parameters can be constructed from the random macroscopic responses obtained through microscale RVE analyses (Chen et al., 2022; Fernandez-Zelaia et al., 2022; Holte et al., 2023; Liu et al., 2021; Liu et al., 2023; Xu et al., 2022a). On the other hand, the variability in macroscopic parameters can also be inversely derived from a large number of dispersed experimental results (Guo et al., 2023; Hossain and Stewart, 2021). However, these studies typically focus on randomness at a single scale (either micro or macro) and have yet to achieve a bottom-up cross-scale modeling approach that integrates microscale stochasticity into full-field macroscopic structural analyses.

In summary, for creep, known as a complex multiscale and multi-mechanism phenomenon, several bottlenecks hinder the realization of a full-field macroscopic creep deformation and failure analysis that incorporates inherent microstructural randomness: (1) The high computational cost in microscale simulations makes it extremely challenging to analyze the effect of random microstructures on the variability in macroscopic material responses; (2) The quantitative relationship between microstructural randomness and the stochastic nature of macroscopic creep constitutive parameters has not yet been established; (3) Full-field creep failure analysis of macroscopic structures, incorporating the stochastic effects of microstructural variability, has yet to be achieved.

To address these challenges, this study focuses on GB-dominated creep failure mechanisms and takes Inconel 617 as a representative material. (1) A novel Monte Carlo (MC) method is established to efficiently obtain the macroscopic creep behavior of RVEs with varying GB characteristics. (2) Building on this, a stochastic dual-scale creep damage model is proposed to quantitatively incorporate the microstructural variability into macroscopic damage evolution. (3) Finally, full-field random creep failure analysis of macroscopic structures is conducted to capture the influence of microstructural randomness. This work bridges the gap between microscale stochasticity and macroscopic creep behavior, providing a robust dual-scale stochastic modeling framework for predicting the creep failure behavior of superalloys and key components. To this end, the remainder of this paper is organized as follows. Firstly, a thorough microscale simulation of creep failure in Inconel 617 is conducted using the CPFE-CZM method. The orientation and area of the GBs are statistically analyzed based on the reconstructed microstructure. The damage behavior of GBs is analyzed, and the relationship between GB damage and the GB characteristics is quantitatively described. Subsequently, as a bridge between microstructure and macroscopic mechanical properties, a MC approach is proposed to derive macroscopic creep deformation and failure behaviors based on the distribution of GB features. The creep rupture time and strain under uniform uniaxial/multiaxial stress state are determined based on the proposed MC approach. Finally, a dual-scale stochastic multiaxial creep damage model is established. The random creep damage and failure behavior of typical structures, influenced by microstructure, is numerically studied based on the established damage model.

## 2. Assumptions and features based on micro-mechanisms

To establish the subsequent model, detailed microscale calculations are first conducted, and model assumptions are proposed based on microscopic failure mechanisms, while quantifying the stochastic characteristics of the microstructure.

### 2.1. Summary of a CPFE-CZM framework for Inconel 617

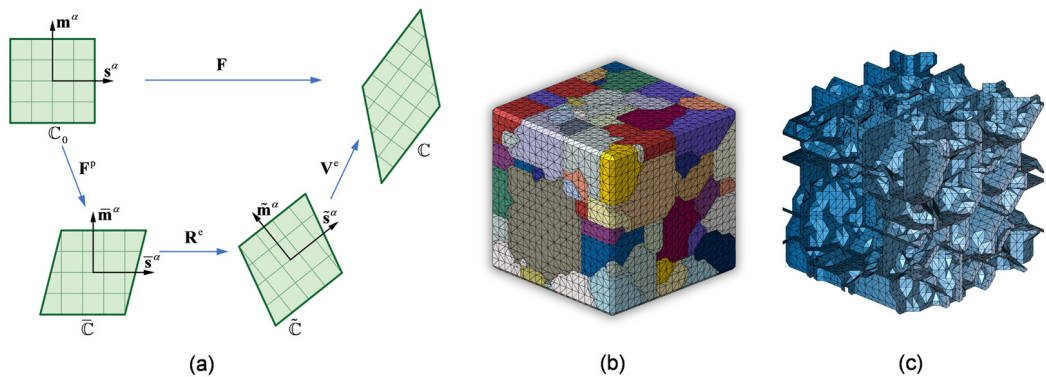

Fig. 2 (a) Kinematics and configurations of single crystals deformation (b) A RVE used in CPFE-CZM simulation (c) GB cohesive elements

To analyze the microscopic creep failure mechanisms of Inconel 617, the CPFE-CZM framework proposed by Zhang and Oskay (2016) and Phan et al. (2017) is used for microscopic simulation. In the CPFE-CZM framework, the dislocation glide and climb are well incorporated in the CPFE model and the GB damage is considered through CZM method. The detailed formulation and parameters of the CPFE-CZM framework is presented in Appendix A. In this CPFE-CZM study, all GBs are assumed to have the same parameters, and the same assumption is used in the current study in both the CPFE-CZM and MC model. In the current work, we focus on leveraging our previously developed CPFE-CZM model, to generate necessary data and develop a MC model to establish the relationship between the GB orientation and area distribution and macroscopic creep damage.

As shown in Fig. 2b, the RVE is established based on the statistical and morphological information obtained from Electron Backscatter Diffraction (EBSD). The software DREAM.3D and parallel polycrystal mesher (PPM) (Cerrone et al., 2014) are used. Random grain orientations have been observed in previous studies (Mo et al., 2013), and is adopted in the current work as well. As shown in Fig. 2c, the zero-thickness cohesive elements are set at the GBs. In this study, the relevant conclusions are extended, and computations are performed using a RVE with an edge length of 320 microns, containing 151 grains.

## 2.2. The creep failure mechanisms of Inconel 617 and assumptions

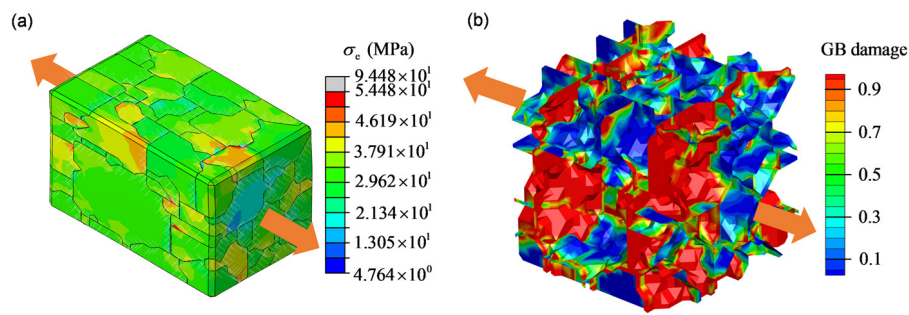

Fig. 3 The CPFE simulation results of (a) von Mises equivalent stress of grains and (b) creep damage of GBs under uniaxial tension $\sigma_e = 28.4$ MPa

Based on the CPFE-CZM framework, the microscopic creep failure process of Inconel 617 at 950 °C can be simulated. As shown in Fig. 3, the deformation of each grain and the damage of each GB are obtained for a RVE under uniaxial tension ($\sigma_e = 28.6$ MPa ).

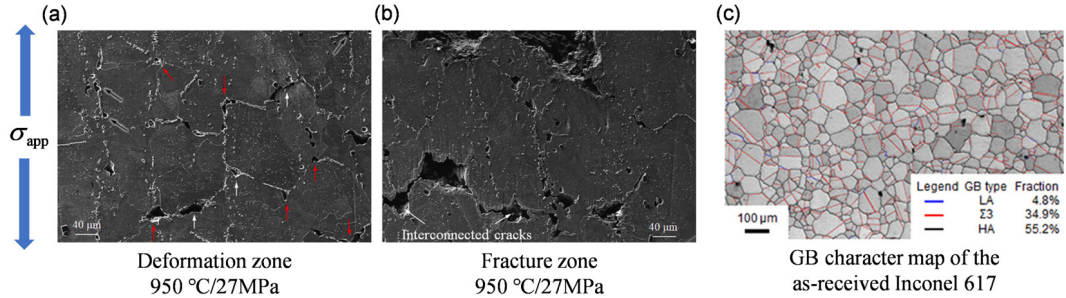

Fig. 4 The SEM images of Inconel 617 of (a) deformation zone and (b) fracture zone (Wang et al., 2021) (c) The GB character map of the as-received Inconel 617 (Wang et al., 2022a)

From experimental observations, the creep failure mechanism of Inconel 617 is related to both temperature and the applied stress level (Kim et al., 2015; Wang et al., 2021). For pressure vessels and pipelines, the main operating conditions involve high temperature, low stress, and long-term creep. In this case, the rupture of Inconel 617 is primarily dominated by GB failure. As shown in Fig. 4 a and b, both in the deformed zone and the fracture zone, there are many creep voids on the GBs perpendicular to the loading direction. There are no significant cracks or voids within the grains leading to final failure of specimen. This scenario also serves as the foundation for many classical GB creep void growth models (Cocks and Ashby, 1982; Sham and Needleman, 1983; Van Der Giessen et al., 1995). In the CPFE-CZM-based simulation results shown in Fig. 3a, the GB failure-dominated creep rupture of the material is well captured.

Therefore, this leads to Assumption I: the long-term creep considered in this study occurs under high temperature and low stress conditions, where the dominant mechanism of material creep rupture is GB failure. In the subsequent modeling, the focus will be on the damage and failure behavior of the GBs.

The GB character map of the as-received Inconel 617 is shown in Fig. 4c. The proportion of low-angle grain boundaries (LAGBs) is relatively small, while the proportion of high-angle grain boundaries (HAGBs) is relatively high. Hence, the HAGBs are relatively significant. Overall, due to the isotropic characteristics of the material, the directional distribution of GBs shows no significant preference. Therefore, it can be considered that the orientation of the GBs is random. Both the EBSD and the GB characteristics map represent the two-dimensional characteristics of GBs, and the inversion of these into three-dimensional features is a challenge. Hence, the reconstructed RVE shown in Fig. 2b is thought as a representative three-dimensional microstructure.

To simplify the subsequent modeling, Assumption II regarding the GB characteristics is proposed: Due to material isotropy, it is considered in this study that the orientation of GBs is randomly distributed with equal probability in all directions. The characteristics of the GBs in the reconstructed three-dimensional microstructure are statistically representative.

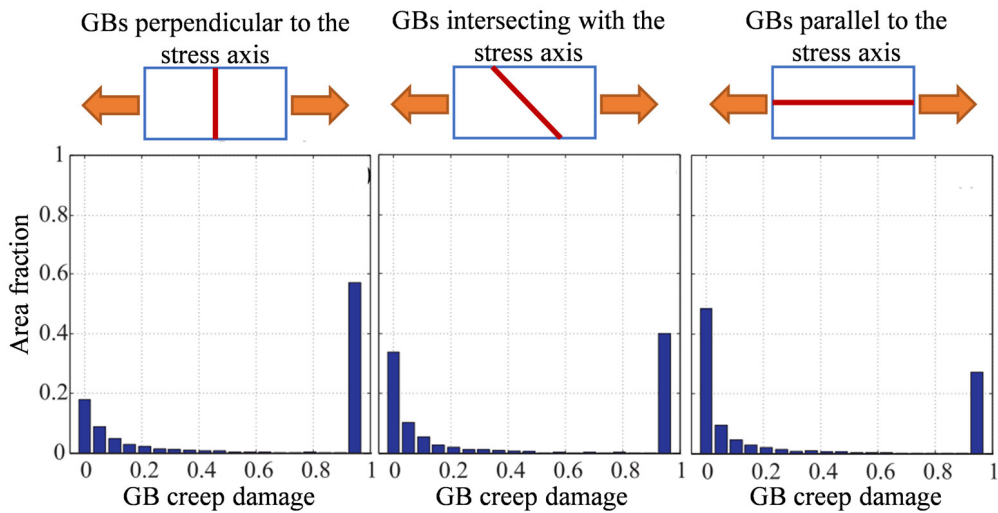

Fig. 5 Creep damage distribution of cohesive elements with different directions with 30% creep strain

(Phan et al., 2017)

The main factors influencing the GB damage process include the GB misorientation and the stress component applied to the GB by the macroscopic stress. As revealed by Phan et al. (2017), creep cracks initially originate at GBs with larger misorientations (i.e., HAGBs) and then propagate to adjacent GBs. However, from a statistical perspective, the primary factor influencing GB damage is the GB orientation. As shown in Fig. 5, the statistical distribution of damage of GBs with different orientations is presented, where the creep strain is 30%. Statistically, the area fraction of GBs with a damage value of 1 is closely related to the GB orientation. The damage is largest on GBs that are nearly perpendicular to the creep stress axis, and smallest on those that are nearly parallel to the creep stress axis. GB damage in other orientations falls in between.

This leads to Assumption III regarding GB damage: statistically, the primary factor influencing GB damage is the GB orientation. The effect of misorientation on the statistical characteristics of GB damage is minimal and is neglected in the model for the sake of simplification.

In addition, regarding the influence of precipitates, Inconel 617 is a typical solid-solution strengthened material, and the proportion of its precipitation-strengthened phases is relatively low (Cabibbo et al., 2008; Zhang et al., 2025). Since this study focuses on the creep rupture caused by GBs, the influence of precipitates on creep is primarily due to the carbide precipitates at the GBs (Kim et al., 2015). Currently, the research on the evolution of carbide precipitates is not clear. The factors influencing the distribution and evolution of carbides include temperature (Cabibbo et al., 2008), misorientation (Salvini et al., 2024), heat treatment (Zhang et al., 2025), etc. Therefore, considering the influence of carbides in a micro-mechanical model, such as CZM used in this study, is challenging. Therefore, Assumption IV is that the evolution of carbide precipitates and their effects are not considered independently at this stage in the modeling of GBs.

## 2.3. Statistical characteristics of GB orientation and area

According to Assumption I, the stochastic nature of the GB distribution plays an important role in the variability of the macroscopic response induced by the microstructure. This study will primarily focus on the geometric characteristics of GBs, specifically GB orientation and area, and their influence on creep failure. The effective normal stress applied on the GB, serving as the key factor controlling GB damage (Cocks and Ashby, 1982), is governed by GB orientation and area. Therefore, the statistical characteristics of GB orientations and areas are required to investigate their influence on creep failure.

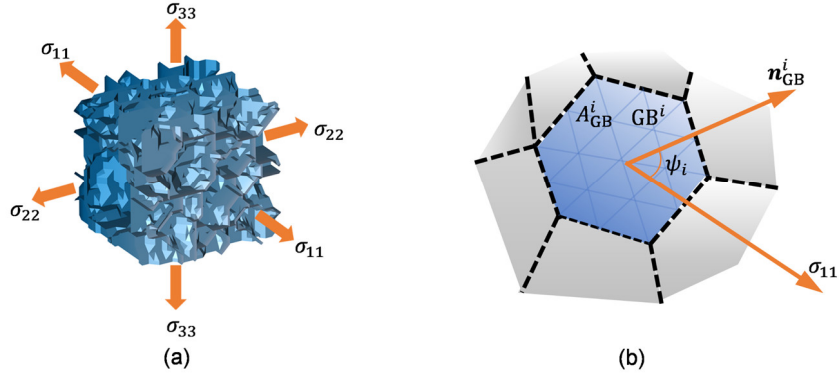

Fig. 6 (a) The configuration of GBs (b) one single GB element set (the blue surface) and its features

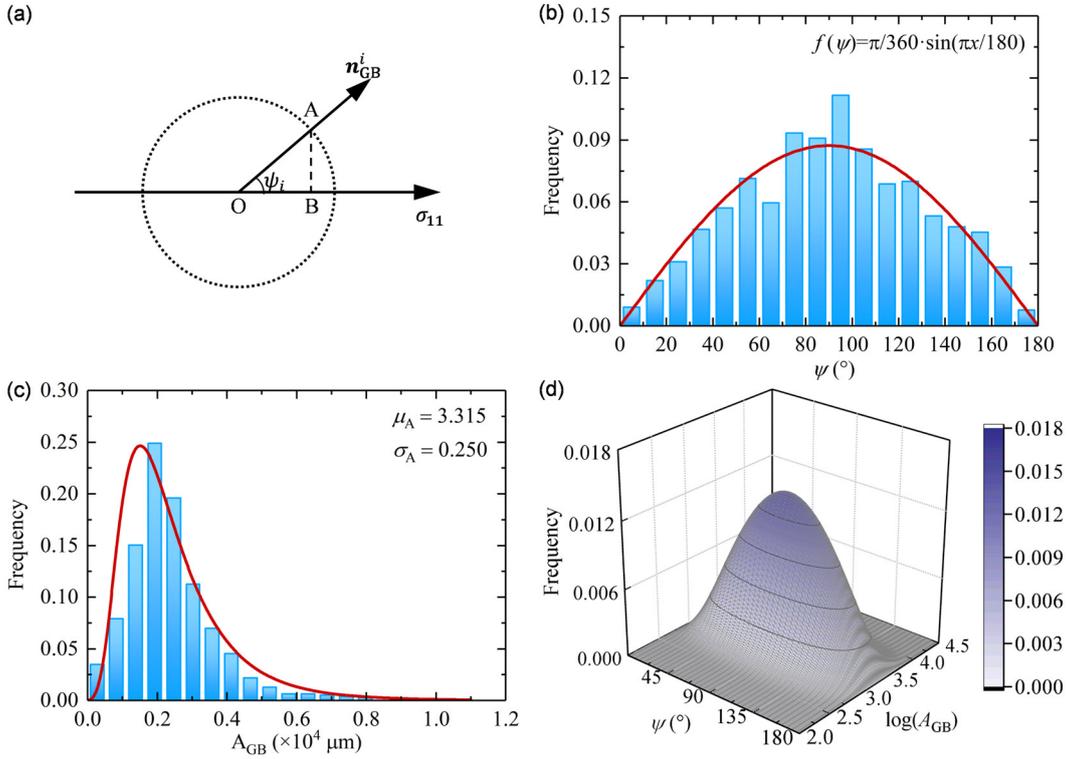

Fig. 7 (a) Schematic diagram of grain boundary orientations $\psi_i$ (b) The distribution of GB element set orientation angle $\psi_i$ (c) The distribution of GB element set area $A_{GB}^i$ (d) Two-dimensional distribution concerning $\psi_i$ and $A_{GB}^i$

As shown in Fig. 2c, the ideal GB configuration is reconstructed when generating grains in the RVE. As shown in Fig. 6a. The RVE is under the stress state with three principal stresses $\sigma_1$, $\sigma_2$ and $\sigma_3$ ($\sigma_1 > \sigma_2 > \sigma_3$). The characteristics of the GBs are statistically analyzed based on GB element sets. As shown in Fig. 6b, a planar GB element set $GB^i$ consists of multiple coplanar GB elements. The orientation of single GB element set $GB^i$ is defined by the angle $\psi_i$ between the normal to the GB element set $\mathbf{n}_{GB}^i$ and the direction of maximum principal stress $\sigma_1$. The area of the GB element set is $A_{GB}^i$.

It is assumed that the orientation of the GBs is distributed isotropically in 3D space. As shown in Fig. 7a, for a given orientation angle $\psi_i$, there exists a corresponding circular ring in 3D space with radius $AB$ and center at point $B$. The circumference of the circular ring is $2\pi AB$. Considering the isotropy of the orientation, the probability density of the angle $\psi_i$ is proportional to $\sin(\psi_i)$. Therefore, the probability density function of the angle $\psi_i$ can be considered to be in the form of a sine function:

$$f(\psi) = \frac{\pi}{360}\sin\left(\frac{\pi\psi}{180}\right) \tag{1}$$

in which $\psi$ is in degrees. The integral of the probability density function over the interval $[0°, 180°]$ equals 1, which satisfies the requirement for a probability density function. Considering spatial symmetry and the characteristics of trigonometric functions, Eqn. (1) can also be written as $f(\psi) = \pi/180 \sin(\pi\psi/180)$ within the range $[0°, 90°]$.

As shown in Fig. 7b, the histogram represents the distribution of GB orientation angles reconstructed from the grain size distribution (Phan et al., 2017), with the edge length of the reconstructed RVE being $500\ \mu m$. Overall, the reconstructed orientation angles $\psi$ align well with the probability density function given in Eqn. (1). The mechanical properties of Inconel 617 used in this study are from Idaho National Labs (Wright et al., 2014). In Fig. 2b and c, the reconstructed RVE contains a total of 908 GB element sets, a number that can reflect the statistical distribution of the GB element sets characteristic sufficiently (Phan et al., 2017). The reconstructed GB areas are shown in Fig. 7c. Assuming that the distribution of GB areas follows a log-normal distribution and fitting the data, it is obtained:

$$A_{GB} \sim \log N(\mu_A = 3.315, \sigma_A = 0.250) \tag{2}$$

in which $A_{GB}$ is measured in μm². Hence, a probability density function of 2D distribution of $\psi \sim \log(A_{GB})$ is obtained:

$$f(\psi, \log(A_{GB})) = \left[\frac{\pi}{360}\sin\left(\frac{\pi\psi}{180}\right)\right] \cdot \left[\frac{1}{\sigma_A\sqrt{2\pi}}\exp\left(-\frac{(y-\mu_A)^2}{2\sigma_A^2}\right)\right] \quad (3)$$

The 2D probability density function is shown in Fig. 7d. The number of containing GBs $N_{GB}$ is determined with a given RVE size represented by RVE edge length $a$. Based on the 2D probability density function, a random sampling can be performed to obtain the 2D distribution of GB characteristics.

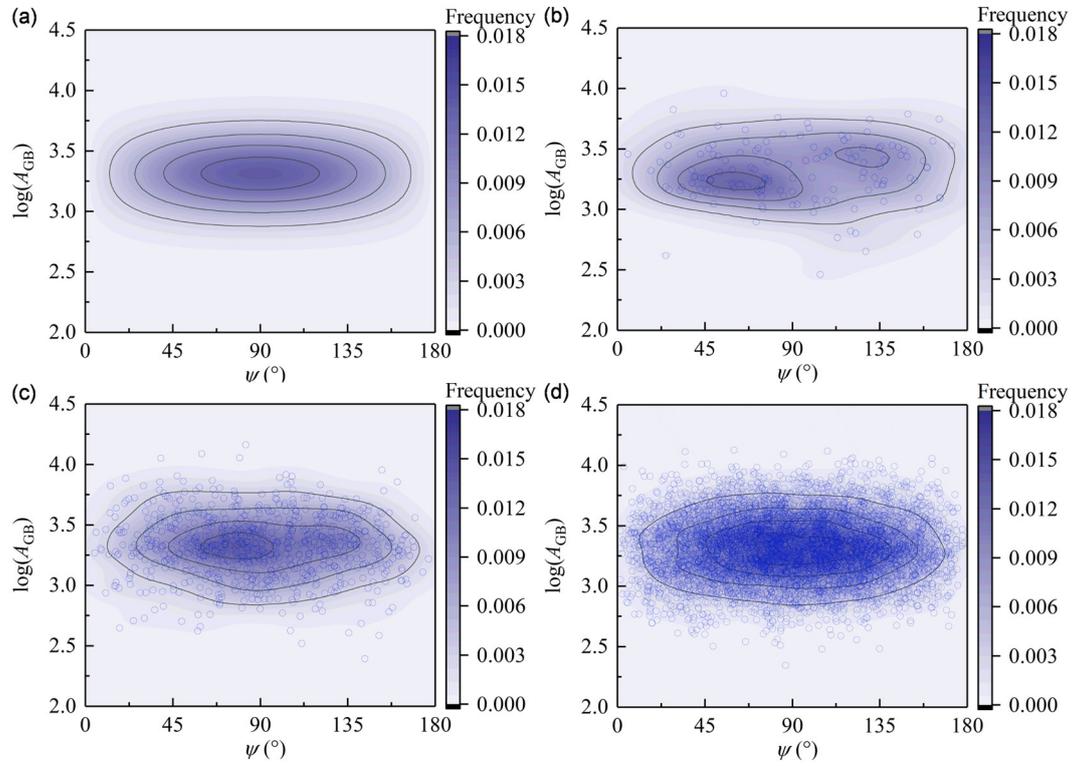

Fig. 8 (a) Two-dimensional distribution concerning $\psi_i$ and $A_{GB}^i$ (b) The Two-dimensional distribution of $\psi_i$ and $A_{GB}^i$ for the RVE with edge length $a = 240$ μm (c) The Two-dimensional distribution of $\psi_i$ and $A_{GB}^i$ for the RVE with edge length $a = 400$ μm (d) The Two-dimensional distribution of $\psi_i$ and $A_{GB}^i$ for the RVE with edge length $a = 800$ μm

The 2D distribution map described in Eqn. (3) is shown in Fig. 8a. Due to the lack of direct GB orientation and area distribution data from EBSD for the material, the comparison can only be made with the GB characteristics of other reconstructed RVEs. The 2D kernel density distribution maps of GB orientation and area from reconstructed

RVEs with edge lengths of 240, 400, and 800 μm are presented in Fig. 8b, c, and d, respectively. It can be observed that as the RVE size increases, more GBs are considered, leading to a better agreement of the 2D kernel density distribution with Eqn. (3). In practice, if the GB characteristics distribution can be directly obtained from EBSD scans, these experimentally derived values can be directly used for subsequent modeling and reconstruction.

## 3. A MC approach based on GB creep damage

Since GB damage is key to characterize creep failure, this section will develop a MC approach based on GB damage. This MC approach will facilitate the input of stress states and random GB distributions to obtain the macroscopic creep damage of the RVE, ultimately yielding the creep curve.

### 3.1. GB creep damage evolution

Firstly, a GB creep damage evaluation function is established. For a RVE considered, it is under the stress state with three principal stresses:

$$\mathbf{\Sigma} = \begin{bmatrix} \sigma_1 & 0 & 0 \\ 0 & \sigma_2 & 0 \\ 0 & 0 & \sigma_3 \end{bmatrix} \tag{4}$$

in which $\sigma_1 > \sigma_2 > \sigma_3$.

For high-temperature, low-stress long-term creep conditions, Inconel 617 exhibits brittle failure, with macroscopic fracture surface perpendicular to the loading direction. In microscale, creep void primarily occurs at GBs perpendicular to the direction of maximum principal stress. It indicates that the creep damage of GBs is dominated by the maximum principal stress. To simplify the issue, this study averages the second and third principal stresses, reducing it to a quasi-two-dimensional stress state:

$$\bar{\mathbf{\Sigma}} = \begin{bmatrix} \sigma_1 & 0 \\ 0 & k \cdot \sigma_1 \end{bmatrix} \tag{5}$$

in which $k = 0.5(\sigma_2 + \sigma_3)/\sigma_1$, i.e.,

$$k = \frac{3\sigma_m - \sigma_1}{2\sigma_1} \qquad (6)$$

This simplification is based on the dominant influence of the maximum principal stress and the relative insignificance of the other principal stresses. For a planer GB element set shown in Fig. 6b, the applied effective normal stress can be calculated as:

$$\sigma_n^i = \mathbf{n}_{GB}^i \cdot \overline{\Sigma} \cdot \mathbf{n}_{GB}^i = \sigma_1 (\cos\psi_i)^2 + k\sigma_1 (\sin\psi_i)^2 \qquad (7)$$

i.e.,

$$\sigma_n^i = \sigma_1 \left[ \left(1 - \frac{3\sigma_m - \sigma_1}{2\sigma_1}\right)(\cos\psi_i)^2 + \frac{3\sigma_m - \sigma_1}{2\sigma_1} \right] \qquad (8)$$

According to Section 2.2 Assumption III, it can be considered that the normal stress acting on the GB element set controls the creep damage evolution. It is noted that the simplification makes the three-dimensional stress state be a quasi-two-dimensional stress state. This simplified approach results in a more concise expression and fewer characteristic parameters. However, it should be noted that the macro multiaxial stress state is considered in Eqn. (8). Hence, a semi-phenomenological power-law equation with certain physical significance is established to incorporate both the effect of stress level, stress state and GB orientation:

$$D_{GB}^i = B \cdot \sigma_1^q \cdot \left[ \left(1 - \frac{3\sigma_m - \sigma_1}{2\sigma_1}\right)\cos^2\psi_i + \frac{3\sigma_m - \sigma_1}{2\sigma_1} \right] \cdot t^p \qquad (9)$$

where $B$ is a material parameter. $q$ and $p$ are stress and time exponents, respectively.

However, it should be noted that the creep damage and failure occur even though the applied normal stress is equal to zero (e.g., $k=0, \psi_i=0$). For example, for uniaxial tension condition, there is apparent creep damage at the GB elements parallel to the maximum principal stress as shown in Fig. 3b. There are two reasons that contribute to this phenomenon: 1) Due to the anisotropy of each grains, the actual local stresses on these GB elements are not zero even though the macroscopic applied stress $\overline{\Sigma}$ results in a zero normal stress component $\sigma_n^i$; 2) As creep evolves, intergranular cracks will propagate from other GBs to those parallel to the direction of maximum principal stress (Phan et al., 2017). In other words, the interaction of GBs will contribute to the failure behavior of RVEs.

To account for this phenomenon, the damage evolution of the GB element sets is proposed to be:

$$\begin{aligned} D_{GB}^i &= B \cdot \sigma_1^q \cdot \exp\left\{\beta\left[\left(1 - \frac{3\sigma_m - \sigma_1}{2\sigma_1}\right)\cos^2\psi_i + \frac{3\sigma_m - \sigma_1}{2\sigma_1}\right]\right\} \cdot t^p \\ &= B \cdot \sigma_1^q \cdot \exp\left\{\beta\left[(1-k)\cos^2\psi_i + k\right]\right\} \cdot t^p \end{aligned} \quad (10)$$

Here, an exponential function is introduced to ensure that creep damage may still occur even when the applied normal stress component $\sigma_n^i$ is zero. $\beta$ is a parameter which reflect the strength of interaction between GBs. According to Eqn. (10), the GB damage evaluation is related to the maximum principal stress and mean stress of the multiaxial stress state.

Physically, the GB damage evolution is highly dependent on the GB creep void growth behavior. As analyzed by Cocks and Ashby (1982), The GB creep growth is a very complicated phenomenon dominated by multiple mechanisms, including boundary diffusion, surface diffusion, power-law creep and their couplings. These mechanisms can be approximated and modeled to establish the relationship between void growth, creep damage, and creep strain. The damage model based on diffusion and void growth can offer physical insight into macroscopic damage and effectively quantify the damage evolution.

Moreover, another important mechanism is GB sliding (Cocks and Ashby, 1980; Raj and Ashby, 1971). GB sliding promotes cavity growth, and at the same time, it induces stress concentration at the GB, significantly affecting GB void growth and the associated damage evolution. GB sliding is influenced by the GB shape, GB diffusion, and precipitate particles (Raj and Ashby, 1971). Mechanistically, void growth, diffusion, GB sliding, and the stress concentration jointly influence the deformation and failure behavior of the GB. These mechanisms generally interact with each other. Moreover, at the microscopic level, factors such as GB misorientation, the evolution of precipitates surrounding the GBs, and the 3D propagation of intergranular cracks more complexly influence GB failure. To date, these mechanisms have not been fully and analytically investigated. It is evident that theoretically considering and incorporating these mechanisms into the GB damage model is highly challenging.

In fact, the proposed GB damage evolution model, Eqn. (10), is a semi-phenomenological model with certain physical significance. The model considers the dominant role of normal stress component on the GB, which is a key factor driving creep void growth by diffusion, thus providing it with certain physical significance. The model is also based on CPFE-CZM calculation results. Thus, complex factors such as stress concentration at the GB, sliding, and 3D propagation of intergranular cracks captured by CPFE-CZM calculation are not physically considered but are introduced into the model in a phenomenological manner.

It is an essential path to develop a GB damage model with stronger physical solvability by clearly defining damage physics as the area fraction of GB void and integrating a void growth model that accounts for the effects of GB diffusion, surface diffusion, and power-law growth. However, further development of the GB damage model requires experimental and theoretical analysis, along with a deeper consideration of the influence of more detailed microscopic mechanisms. This will be the focus of future research.

With the creep damage evaluation of single GB element set, i.e., Eqn. (10), an effective GB damage of RVE can be defined. Since the intergranular creep crack mainly occur at the GB perpendicular to the maximum principal stress (see Fig. 3b), the effective area of GB is defined as:

$$\overline{A}_{GB}^{i} = A_{GB}^{i} \cdot \left| \cos \psi_i \right| \tag{11}$$

The total effective area of the RVE is expressed as:

$$A_{RVE}^{eff} = \sum_{i=1}^{N} \overline{A}_{GB}^{i} \tag{12}$$

Furthermore, the effective GB damage of the RVE is defined as:

$$D_{\text{GB}}^{\text{eff}} = \sum_{i=1}^{N} D_{\text{GB}}^{i} \frac{\overline{A}_{\text{GB}}^{i}}{A_{\text{RVE}}^{\text{eff}}} \quad (13)$$

With the simulation results of GB damage versus creep time, the effective damage of RVE versus creep time is also obtained according to Eqn. (13). It is noted that the damage of each GB $D_{\text{GB}}^{i}$ represents the degeneration of GB caused by creep void. When $D_{\text{GB}}^{i}$ reaches 1, it indicates that the void has evolved to a critical size. However, since this semi-phenomenological damage model does not explicitly model the relationship between cavity growth and damage, it cannot explicitly determine the critical cavity size. The effective GB damage of RVE $D_{\text{GB}}^{\text{eff}}$ is defined as the weighted summation of $D_{\text{GB}}^{i}$. The weighting strategy is based on the damage area projecting onto the direction perpendicular to the maximum principal stress. This further emphasizes the importance of the maximum principal stress in the process of long-term creep, which shows a quasi-brittle failure mechanism.

### 3.2. Stochastic creep behavior and damage evolution of RVE

To describe the macroscopic creep behavior of the RVE, it is necessary to introduce appropriate macroscopic creep and damage models. Due to the apparent primary and secondary creep of Inconel 617 from experimental research (Wright et al., 2014), the time-hardening creep model is adopted:

$$\varepsilon_{\text{eq}} = A \cdot \sigma_{\text{e}}^{n} \cdot t^{m} \quad (14)$$

in which $A$ is the amplitude parameter, $n$ and $m$ are the stress exponent and time exponent, respectively. $\varepsilon_{\text{eq}}$ is the macro equivalent strain expressed as:

$$\varepsilon_{\text{eq}} = \sqrt{\frac{2}{3}\left(\varepsilon_{ij} - \frac{\varepsilon_{kk}}{3}\delta_{ij}\right)^{2}} \quad (15)$$

And $\sigma_{\text{e}}$ is the macroscopic von Mises equivalent stress applied to the RVE:

$$\sigma_{\text{e}} = \sqrt{\frac{3}{2}s_{ij}s_{ij}} \quad (16)$$

in which

$$s_{ij} = -p\delta_{ij} + \sigma_{ij} \tag{17}$$

$$p = \sigma_m = \frac{\sigma_{kk}}{3} = \frac{\sigma_1 + \sigma_2 + \sigma_3}{3} \tag{18}$$

The multiaxial creep strain rate can be expressed as:

$$\dot{\varepsilon}_{ij} = \frac{3}{2} A \cdot m \cdot \sigma_e^{n-1} s_{ij} \cdot t^{m-1} \tag{19}$$

The macro creep damage $\omega$ is introduced into the creep strain as:

$$\dot{\varepsilon}_{ij} = \frac{3}{2} A \cdot m \cdot \left(\frac{\sigma_e}{1-\omega}\right)^{n-1} \frac{s_{ij}}{1-\omega} \cdot t^{m-1} \tag{20}$$

which can be degenerated into the uniaxial form as:

$$\dot{\varepsilon}_{eq} = A \cdot m \cdot \left(\frac{\sigma_e}{1-\omega}\right)^n \cdot t^{m-1} \tag{21}$$

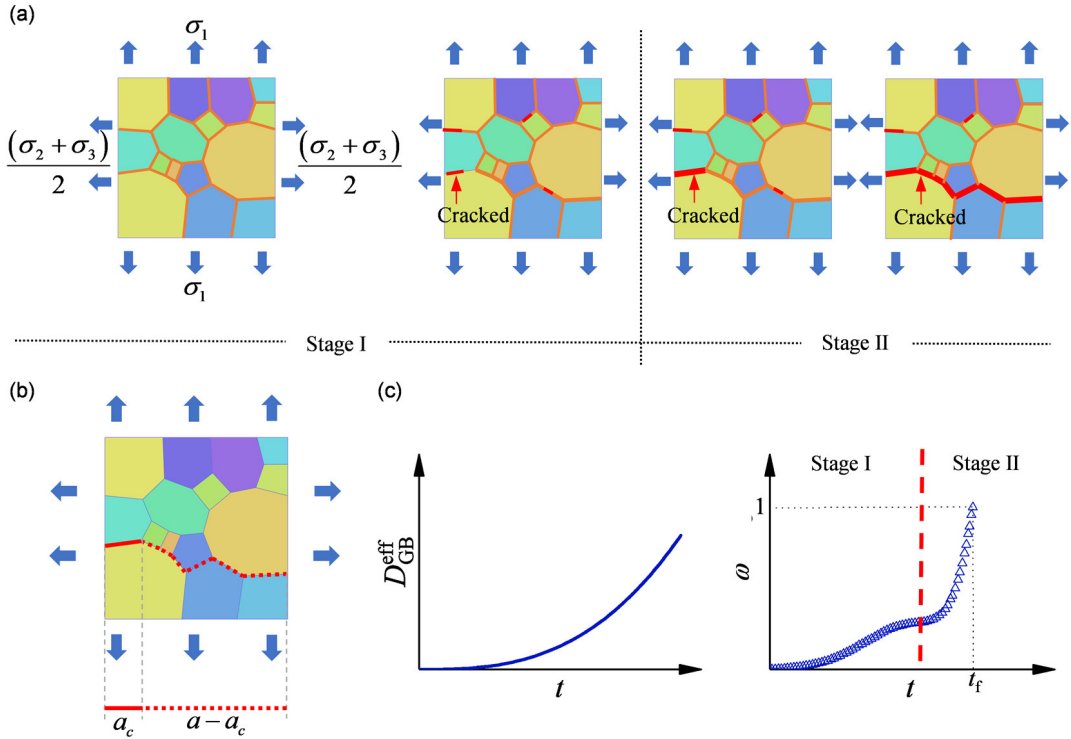

Fig. 9 (a) The GB damage evolution from global GB damage accumulation (Stage I) to local GB crack growth (Stage II) (b) The physical definition of macro creep damage $\omega$ (c) The curves of effective GB damage and macro creep damage

According to Eqn. (13) and Eqn. (21), two different kinds of creep damage are used to describe the creep damage of RVE. As shown in Fig. 9a, the damage accumulation and evolution process of GBs is divided into two stages. In stage I, GB damage occurs globally throughout the RVE. The orange-colored GBs undergo uniform damage accumulation. Then, some micro-cracks initiates in stage I. Therefore, $D_{\text{GB}}^{\text{eff}}(t)$ is used to describe the global GB damage level of the RVE. In stage II, a dominant main GB crack (the red-colored GBs) initiates and propagates. When this crack penetrates through, the RVE is considered to have failed. Hence, the macro creep damage $\omega(t)$ is defined as:

$$\omega(t) = \frac{a_c}{a} \tag{22}$$

in which $a_c$ and $a$ are the effective cracked length and edge length of RVE as shown in Fig. 9b. This definition also corresponds to the net stress $\sigma/(1-\omega)$ in Eqns. (20) and (21). To avoid the singularity that may occur when $\omega$ approaches 1, it is assumed that creep failure occurs when $\omega$ reaches 0.99 in both the model calibration and MC calculations. This rule is also applied in the subsequent macroscopic FEM calculations. As shown in Fig. 9c, the accumulation of GB damage occurs smoothly, even with the presence of the main crack. However, the accumulation of macro damage differs significantly between stage I and stage II. In stage I, due to the initiation of GB micro-cracks, macroscopic damage $\omega(t)$ accumulates slowly. In stage II, with the initiation and propagation of the main GB crack, macroscopic damage rapidly increases until it reaches unity.

The whole uniaxial tensile creep curves with primary, secondary and tertiary stages can be obtained from CPFE-CZE calculation. Hence, the GB effective damage $D_{\text{GB}}^{\text{eff}}(t)$ and macro damage $\omega(t)$ can be obtained from CPFE-CZE calculation. As shown in Fig. 10a, the uniaxial creep strain curve (the red symbols) is obtained through the CPFE-CZM calculation. The creep damage is reflected in the obtained creep strain curve. The creep strain rate $\dot{\varepsilon}_{eq}$ can be obtained through a numerical difference calculation according to the creep strain curve. Hence, combining with Eqn. (21), the macroscopic creep damage $\omega(t)$ is obtained from the CPFE-CZM calculation of RVE.

It is noted that the definition of macroscopic creep damage $\omega(t)$ is based on the macroscopic creep model and damage model. In present work, the time-hardening creep model Eqn. (19) and Kachanov-Rabotnov type damage model is used. Meanwhile, as shown in Fig. 10b, the effective GB creep damage $D_{\text{GB}}^{\text{eff}}(t)$ can be obtained according to Eqn. (13). Finally, the relation between macroscopic creep damage $\omega(t)$ and the effective GB creep damage $D_{\text{GB}}^{\text{eff}}(t)$ is established as shown in Fig. 10c. To describe the rapid accumulation of macroscopic damage in stage II, an inverse hyperbolic tangent function form is assumed to represent the relationship between macro and micro damage:

$$\omega = a \cdot \text{arctanh}\left(b \cdot D_{\text{GB}}^{\text{eff}}\right) \tag{23}$$

It is noted that both $D_{\text{GB}}^{\text{eff}}(t)$ and $\omega(t)$ describe the degradation of the load-bearing capacity of material, but they are defined differently and have distinct physical meanings at different scales. In microscale, $D_{\text{GB}}^{\text{eff}}(t)$ is used to describe the global GB damage level of the RVE. In macroscale, macroscopic creep damage $\omega(t)$ physically represents the degradation of load-bearing capacity caused by the propagation of macroscopic creep cracks. Both types of damage describe the phenomenological relationship between these two types of damage as presented in Eqn. (23). In the following sections, such a phenomenological formula is found to be valid.

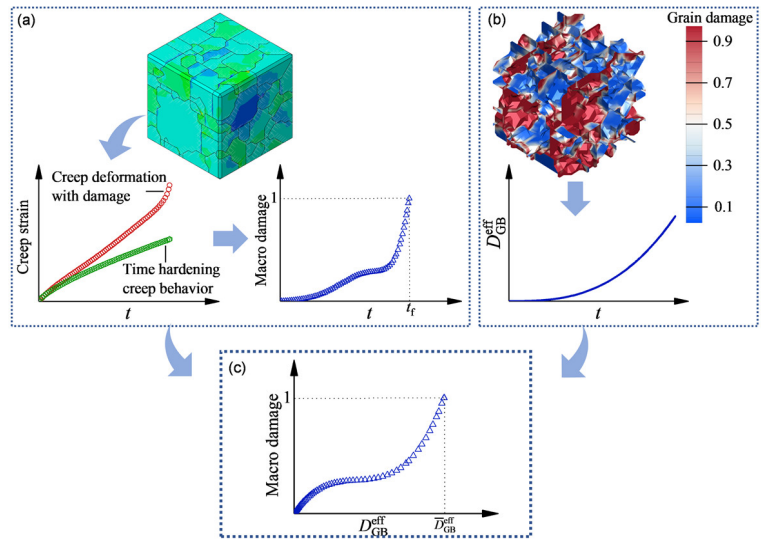

Fig. 10 The process to establish the relation between the macroscopic damage and effective GB damage (a) obtaining the macroscopic creep damage from RVE response (b) obtaining the effective GB creep damage from GB cohesive element (c) the relation between the macroscopic damage and effective GB damage

It should be noted that the macroscopic creep deformation and damage model described in this section is only used to describe the macroscopic creep response of the RVE. It does not account for the local responses at each grain and GB level. However, this macroscopic model establishes the relationship between the random GB characteristics within the RVE and the macroscopic creep damage process through the correlation between the two types of damage $D_{\mathrm{GB}}^{\mathrm{eff}}(t)$ and $\omega(t)$ described above.

## 3.3. Creep failure calculation using MC method

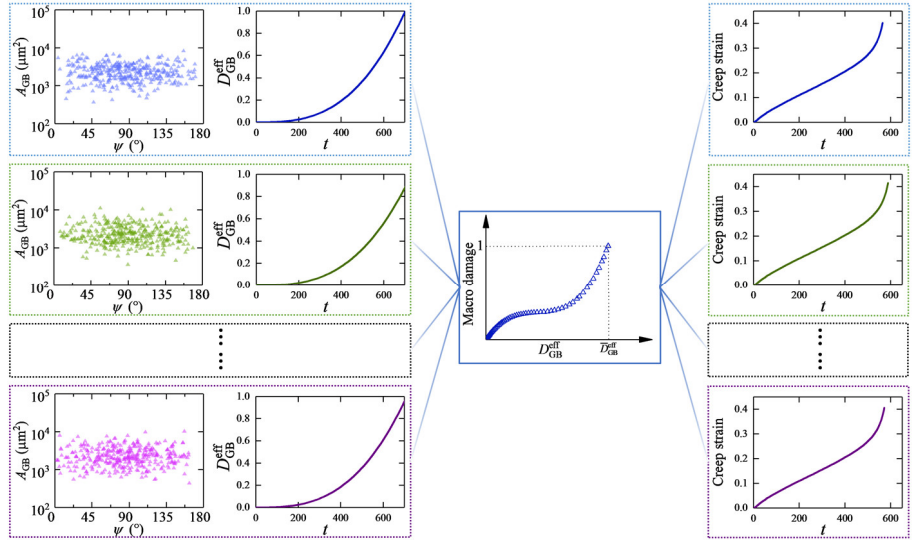

Fig. 11 The MC approach from GB distribution characteristics to creep curves

Based on the relation between the macroscopic creep damage $\omega(t)$ and the effective GB creep damage $D_{\mathrm{GB}}^{\mathrm{eff}}(t)$, a MC approach can be obtained from the GB distribution characteristics to creep curves. As shown in Fig. 11, for a given number of GBs, sets of GBs that conform to the GB characteristics distribution in Section 2.3 are generated. For each set of GBs, the effective GB creep damage $D_{\mathrm{GB}}^{\mathrm{eff}}(t)$ can be obtained according to Eqns. (11)~(13). Then, based on the relation between $\omega(t)$ and $D_{\text{GB}}^{\text{eff}}(t)$, the creep curves corresponding to these sets of GBs can be achieved according to Eqn. (20). Finally, the MC method establishes the relationship between the randomly distributed GB characteristics and the corresponding macroscopic creep response through a series of assumptions and complex modeling.

Moreover, the MC approach is also expected to be effective for multiaxial stress state. The reason is that the multiaxial stress state is considered in Eqns. (10) and (20). Hence, the proposed framework enables prediction of the material creep deformation and failure behavior under various macroscopic stress states solely calibrated by the uniaxial tensile experimental data. The detailed verification of cases under different multiaxial stress states are presented in the following sections. It also should be noted that the proposed MC approach can only obtain the macroscopic strain response of RVEs. In the MC approach, the GBs are not explicitly modeled and analyzed. Hence, the propagation of 3D intergranular crack cannot be obtained. The effective damage $D_{\text{GB}}^{\text{eff}}(t)$ can only provide the failure process of the GBs from a statistical perspective. Moreover, the GB characteristics input into the MC approach are initial ones. The changes in GB characteristics due to recrystallization and grain growth during the creep process are not adequately considered. However, this limitation is manageable as recrystallization and grain growth are not significant in the high-temperature for low-stress long-term creep conditions discussed in this study.

The MC calculation process is realized through the software MATLAB using a personal computer. The MC calculation time versus the number of GBs is shown in Fig. 12. When the number of GBs in each set is 1000, the required computation time is approximately 0.1 seconds. However, the CPFE-CZM calculation for an RVE containing 908 GBs shown in Fig. 2b generally exceeds 25 hours using the software Abaqus (A multi-core strategy using 16 cores is employed). Both the MC and the CPFE-CZM calculation are based on the same personal computer platform (Intel Core i9-12900KF, Windows 10). If the goal is merely to obtain macroscopic quantities such as creep curves and rupture times of RVE under specific stress states, rather than microscopic quantities like local stresses, then a calibrated MC model will significantly reduce the required computation time.

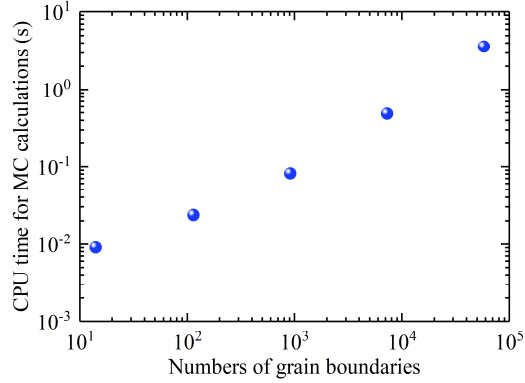

Fig. 12 The MC calculation time versus number of GBs using MATLAB

## 3.4. MC model calibration process

The calibration process of the model is shown in Fig. 13. Firstly, the CPFE-CZM computation with the initial parameters is conducted. And the RVE creep curves and the evolution of GB damage are obtained. Accordingly, the GB damage model Eqn. (10) can be fitted based on the damage evolution of GBs in RVE. Secondly, the macroscopic damage $\omega(t)$ from the RVE creep curves and effective GB damage $D_{\text{GB}}^{\text{eff}}(t)$ based on the CPFE-CZM computational results and Eqn. (13) are derived. Thirdly, the relationship between $\omega(t)$ and $D_{\text{GB}}^{\text{eff}}(t)$, i.e., Eqn. (23), is fitted. Fourthly, the MC calculations to obtain a set of dispersed creep curves based on the random distribution of GB characteristics are performed. Convergent creep strain curves under different applied stress can be obtained and used to represent the macroscopic computational creep behavior of material. Finally, the CZM model parameters based on the differences between the creep curves from the MC calculations and experiments are adjusted until convergence. Therefore, during the model calibration process, only the parameters related to CZM need to be adjusted, most of which have already been determined by Phan et al. (2017). The other parameters are essentially fitted directly from the CPFE-CZM calculation results, which is a relatively simple and reliable forward process. Although the calibration of the MC parameters involves the CPFE-CZM calculation framework, the calibrated MC approach does not require any CPFE-CZM calculations to obtain the macroscopic creep response related to the microstructure.

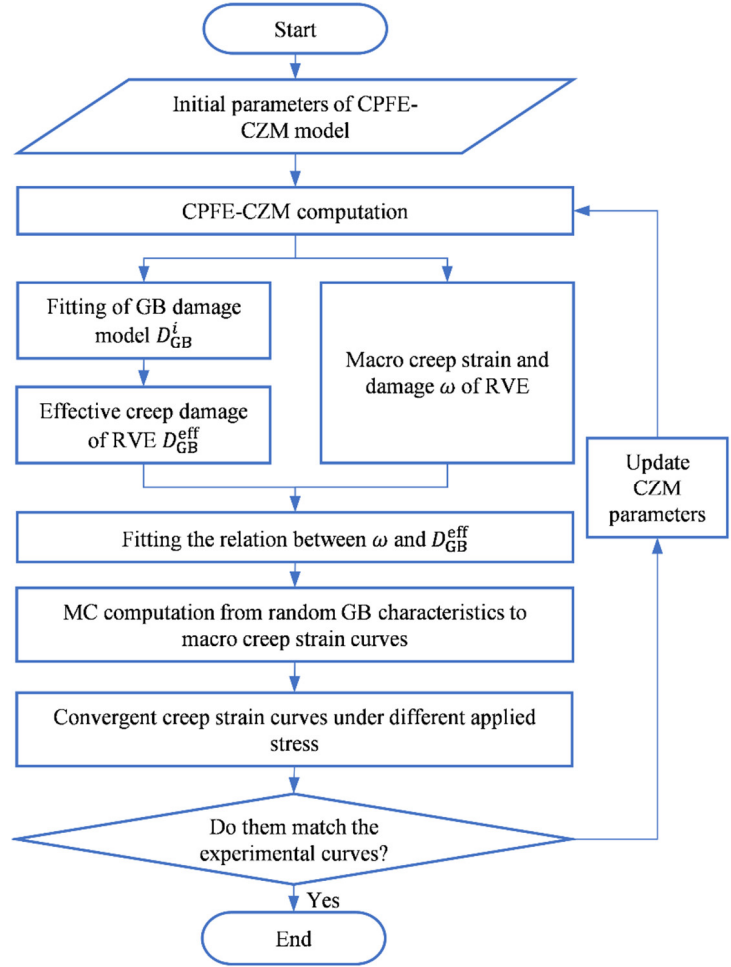

Fig. 13 The calibration process of the parameters involved in MC procedure

A calibration of model parameters for Inconel 617 at 950°C is conducted in Appendix B. The MC results of creep curves under uniaxial tension with $\sigma_e = 28.6$ MPa and $\sigma_e = 18.5$ MPa are shown in Fig. 14. The computational results from the calibrated MC model generally agree with the experimental results used for calibration, particularly regarding the creep rupture time of the specimens. It is noted that the CZM parameters ($\alpha, k_n, \delta_c, r, p, C, T_c$) are based on the prior values obtained by Phan et al. (2017). Meanwhile, the models for creep ($A, m, n$) and damage ($B, \beta, p, q, a, b$) are forward-fitted based on CPFE-CZM calculation results. For a complex multi-parameter model, there are multiple possible calibrated parameters sets, and the goal of calibration is to identify one set of them using a reasonable number of tests to balance accuracy and cost. Then, we will validate the current calibrated values from multiple aspects to demonstrate their good interpolation and extrapolation capabilities.

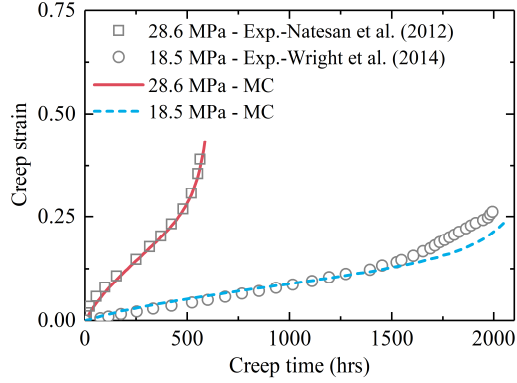

Fig. 14 The comparison of creep curves calculated through the MC approach and experimental data

In the process of obtaining the macroscopic creep deformation of the RVE from microstructural features, the calibrated MC approach is independent surrogate model without any CPFE-CZM computation involved. To validate that the calibrated MC method can serve as a surrogate model for the CPFE-CZM framework, additional RVE configurations and multiaxial stress conditions are systematically tested and compared. The verification process is presented in Appendix C.

## 4. Creep failure results under uniform macro stress states

Based on the proposed MC approach, the equivalent creep strain and creep rupture strain and time under different uniaxial and multiaxial uniform macro stress state can be calculated. In this work, creep rupture signifies that a crack has propagated through the entire RVE. It is noted that the MC approach can only be used under a specific uniform macroscopic stress state $\Sigma$. Moreover, only when the RVE contains a sufficient number of GBs, then the creep damage and rupture behavior of the material at the specimen level can be obtained. Ideally, test results of specimens from the same source should demonstrate good consistency.

## 4.1. Uniaxial creep failure

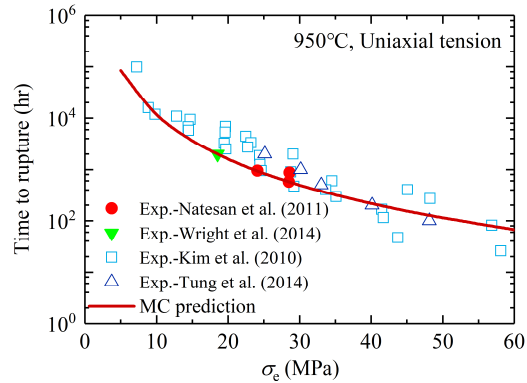

Fig. 15 The MC predicted and experimental creep rupture time from different sources

For creep deformation and failure issues, the most common experimental method is the uniaxial tensile creep test. The calibration in Section 3 is based on experimental results with applied stress of 18.5 MPa and 28.6 MPa. Furthermore, the uniaxial creep rupture time under different applied stress can also be obtained from the MC approach. As shown in Fig. 15, the predicted creep rupture time of Inconel 617 at 950 °C is plotted in red solid line. Only the converged average rupture time predicted by the MC method is provided here. This is because the differences in material initial state, processing history, and testing conditions are not considered in this study. From the perspective of homogenization, the converged average response is representative only when the RVE size is sufficiently large, and it can be used for comparison with macroscopic test results. The creep rupture time decreases with the increasing applied stress. Overall, the predicted values are in agreement with the experimental results.

The experimental data marked by solid symbols are consistent with the material source and testing conditions used for model calibration in this study (sourced from Idaho National Laboratory and tested at Argonne National Laboratory). It should be noted that the experimental data represented by the hollow blue square and triangle symbols come from other material sources (Kim et al., 2010; Tung et al., 2014). This is also the reason for the deviations between the blue square and triangle points and the predicted values. When the material source and testing institution are not the same, uncertainty factors such as fluctuations in testing conditions, initial damage, material properties, and surface conditions will affect the creep behavior. Considering these uncertainties, the creep failure of material will exhibit significant dispersion (Hossain and Stewart, 2021). Nevertheless, the MC predicted results shown in Fig. 15 demonstrate a certain level of accuracy in predicting experimental results from different sources.

It is noted that this study mainly focuses on the creep behavior of Inconel 617 at 950°C for durations under 10,000 hours, in which scenario the GB opening and sliding are considered by the CPFE-CZM framework and the MC model. Due to the lack of experimental data and the expensive cost in simulation, creep beyond 10,000 hours and the associated decrease in activation energy (Maruyama et al., 2022; Maruyama et al., 2017), along with other specific scenarios, were not considered. Overall, the MC predictions are fairly accurate between 300 and 10,000 hours. For high-stress creep situations under 300 hours, the damage mechanism shifts to intragranular void formation (Wang et al., 2021), which reduces the accuracy of the GB degradation-based MC predictions.

## 4.2. Biaxial creep failure

As discussed in Section 3.3, the proposed MC approach can be used to predict the creep behavior and failure under multiaxial macroscopic stress state. As one of the widely existed stress state in pressure vessels and piping at high temperature, the biaxial creep behavior is widely studied (Sanders et al., 2017). The experimental results of biaxial creep failure are obtained using pressurized tubes (Wright and wright, 2013). In the pressurized tube test, the equivalent strain is calculated based on the change of diameter, and the effective stress is calculated based on von Mises equivalent stress (Tung et al., 2014; Wright and wright, 2013). The macroscopic applied stress state is expressed as:

$$\Sigma = \begin{bmatrix} \dfrac{2\sqrt{3}\sigma_e}{3} & 0 & 0 \\ 0 & \dfrac{\sqrt{3}\sigma_e}{3} & 0 \\ 0 & 0 & 0 \end{bmatrix} \tag{24}$$

As shown in Fig. 16a, the equivalent creep strain versus creep time of pressurized tube tests under different effective stress is presented in discrete symbols. The MC predicted results are plotted in lines. It is found that there is a difference between the curve predicted by the MC method and the experimental strain curve obtained from the pressurized tube tests. The creep rupture times of different pressurized tubes are shown in Fig. 16b. The predicted creep rupture time agree well with the experimental results except only one case ( $\sigma_e = 30.3$ MPa ). The reason for the discrepancies in the predicted and experimental curves is that the MC approach can only predict the creep behavior and failure under macroscopic uniform stress states. However, the stress states in the pressurized tube test are not strictly as uniform as described in Eqn. (24). As a structure, the stress state of a pressurized tube is complex and related to the tube's inner diameter, outer diameter, and length.

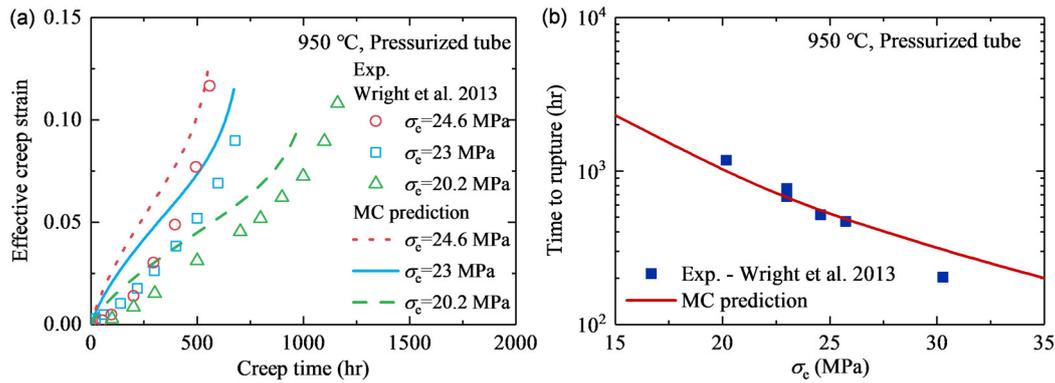

Fig. 16 (a) The MC calculated and experimental biaxial creep curves (b) The MC predicted and experimental biaxial creep rupture time of pressurized tube

### 4.3. Multiaxial creep failure

The multiaxial creep rupture is highly related to the stress state (Cocks and Ashby,

1982; Wen and Tu, 2014; Yatomi and Nikbin, 2014). To describe the multiaxial creep failure, the macro stress triaxiality is introduced:

$$\eta = \frac{\sigma_m}{\sigma_e} \qquad (25)$$

in which $\sigma_m$ and $\sigma_e$ are the hydrostatic stress and von Mises equivalent stress, respectively. Based on the MC calculation results, the equivalent strain versus creep time under different $\sigma_e$ and $\eta$ are shown in Fig. 17. Overall, despite the differences in stress states, these creep curves exhibit similarities in their shapes and trends. With the increase of macro stress triaxiality, the creep rupture time and strain decrease for the cases under same $\sigma_e$.

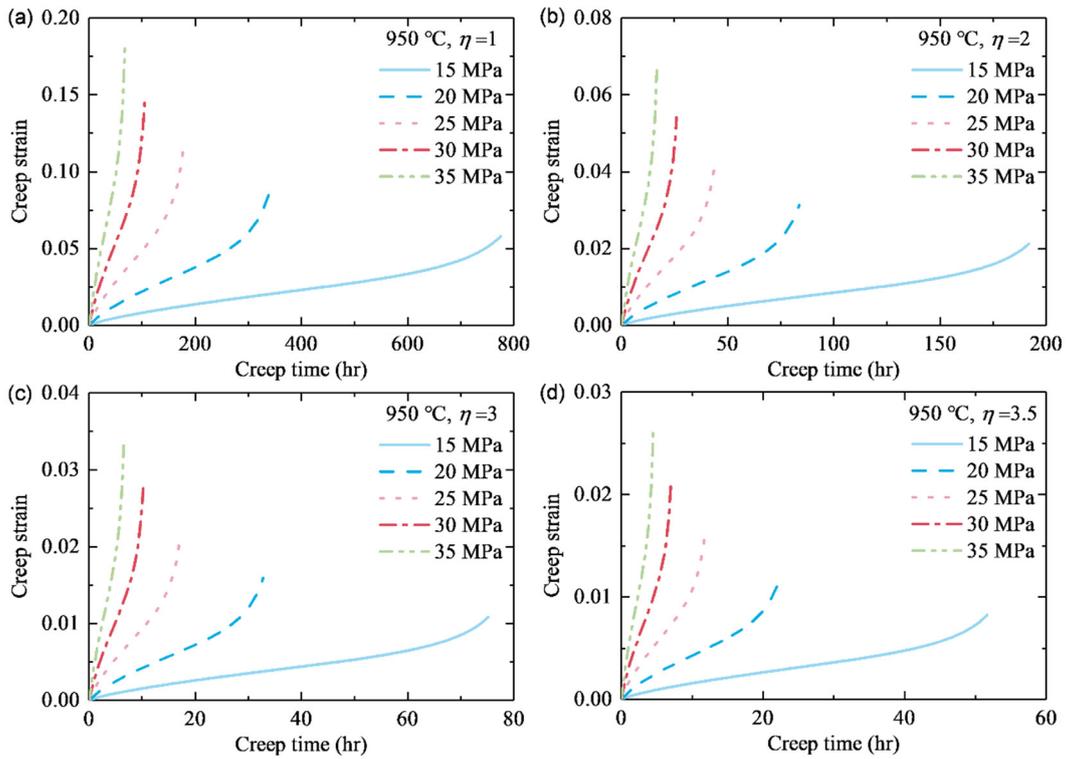

Fig. 17 The curves of multiaxial equivalent creep strain under different macro stress levels and triaxialities

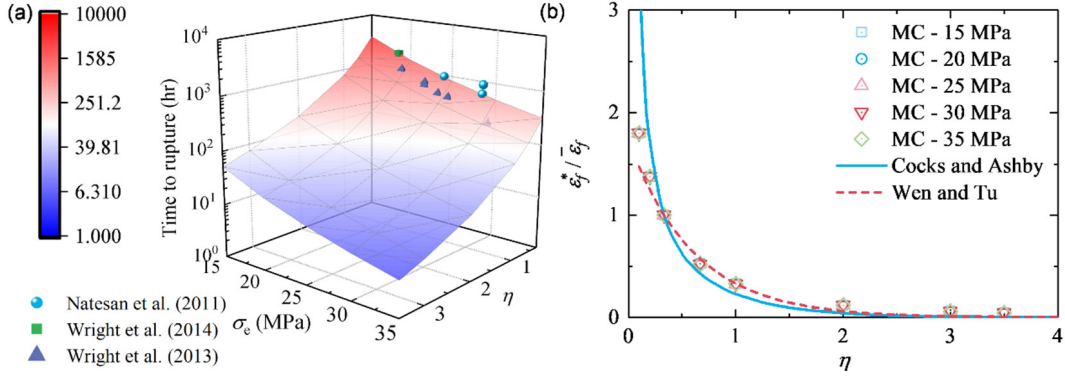

Fig. 18 The multiaxial (a) creep rupture time and (b) creep rupture strain influenced by stress triaxialities $\eta$ and applied stress level $\sigma_e$

As shown in Fig. 18a, the creep rupture time decrease with increasing $\sigma_e$ and $\eta$. Due to the lack of multiaxial test results, the data in Fig. 18a only includes experimental results from uniaxial ($\eta=1/3$) and biaxial ($\eta=\sqrt{3}/3$) stress states (which is also is also a kind of multi-axial stress state). Due to the lack of multiaxial experimental data, the prediction of multiaxial creep rupture time beyond biaxial conditions is incomplete. Therefore, the multiaxial creep rupture strain is used here to validate the accuracy of the MC predictions under multiaxial stress states.

Fig. 18b shows the ratio between multiaxial creep rupture strain $\varepsilon_f^*$ and uniaxial creep rupture strain $\bar{\varepsilon}_f$. Moreover, two practical models describe the relation between the multiaxial and uniaxial rupture strain used in creeping condition are also shown in Fig. 18b:

Cocks-Ashby model (Cocks and Ashby, 1982):

$$\frac{\varepsilon_f^*}{\varepsilon_f} = \sinh\left[\frac{2}{3}\left(\frac{n-0.5}{n+0.5}\right)\right] \Big/ \sinh\left[2\left(\frac{n-0.5}{n+0.5}\right)\frac{\sigma_m}{\sigma_e}\right] \quad (26)$$

Wen-Tu model (Wen and Tu, 2014):

$$\frac{\varepsilon_f^*}{\varepsilon_f} = \exp\left[\frac{2}{3}\left(\frac{n-0.5}{n+0.5}\right)\right] \Big/ \exp\left[2\left(\frac{n-0.5}{n+0.5}\right)\frac{\sigma_m}{\sigma_e}\right] \quad (27)$$

It is found that the MC predicted results consistently fall between the curves corresponding to the two models. This also demonstrates that the MC approach proposed in this paper is effective and reliable for predicting multiaxial creep failure of alloys.

It should be noted that the Cocks-Ashby and Wen-Tu models are both established based on GB void growth behavior. These models describe the effect of hydrostatic stress on GB creep voids and creep ductility. The GB creep damage model Eqn. (10) in this study also effectively incorporates the influence of multiaxial stress. Overall, the evolution model of GB damage in this study effectively describes the phenomenon of accelerated GB damage in polycrystals due to increased hydrostatic stress. Since the Cocks-Ashby and Wen-Tu models have been validated with numerous experimental parameters and proven to be effective (Wen et al., 2016), they can serve as reliable ground truth models. This study merely uses the models to validate the effectiveness of the proposed model in describing multiaxial failure, as these models are widely accepted and applied.

## 5. A dual-scale stochastic multiaxial creep damage model and numerical application

In finite element analysis for creep failure of structures, the use of a sufficiently fine mesh is required to obtain accurate computational results. However, the underlying micro structures are different at different structural elements, leading to the difference of creep behaviors. To reflect this variability in structural calculations, a stochastic creep damage model is developed.

### 5.1. The dual-scale stochastic creep damage model and numerical procedure based on FEM

It should be noted that the stochastic damage behavior of the RVE is influenced by both the size of the RVE and the applied stress state. Specifically, as the RVE size increases and the stress triaxiality rises, the dispersion of creep rupture time and rupture strain obtained from MC calculations decreases. A detailed analysis of this phenomenon is provided in Appendix D. Therefore, it is essential to incorporate both the RVE size and the stress state into the dual-scale stochastic creep damage model.

According to the GB damage evaluation Eqn. (10), the influence of stress state on the damage rate is primarily reflected in the middle part of the formula, i.e., $\exp\{\beta[(1-k)\cos^2\psi_i + k]\}$. Hence, an effective creep damage rate factor $K$ is proposed:

$$K = \sum_{i=1}^{N} \exp\{\beta[(1-k)\cos^2\psi_i + k]\} \frac{\overline{A}_{GB}^i}{A_{RVE}^{eff}} \tag{28}$$

in which the influence of stress state is reflected by the change of $k$, and the effect of RVE size is reflected by the number $N$ of the considered GBs. It is noted that, in the proposed dual-scale stochastic creep damage model, $K$ is the only random variable that is influenced by microstructural variability. The probability histograms of $K$ with different RVE sizes under different stress states are shown in Fig. 19. In all cases, the distribution of $K$ essentially follows a Gaussian distribution.

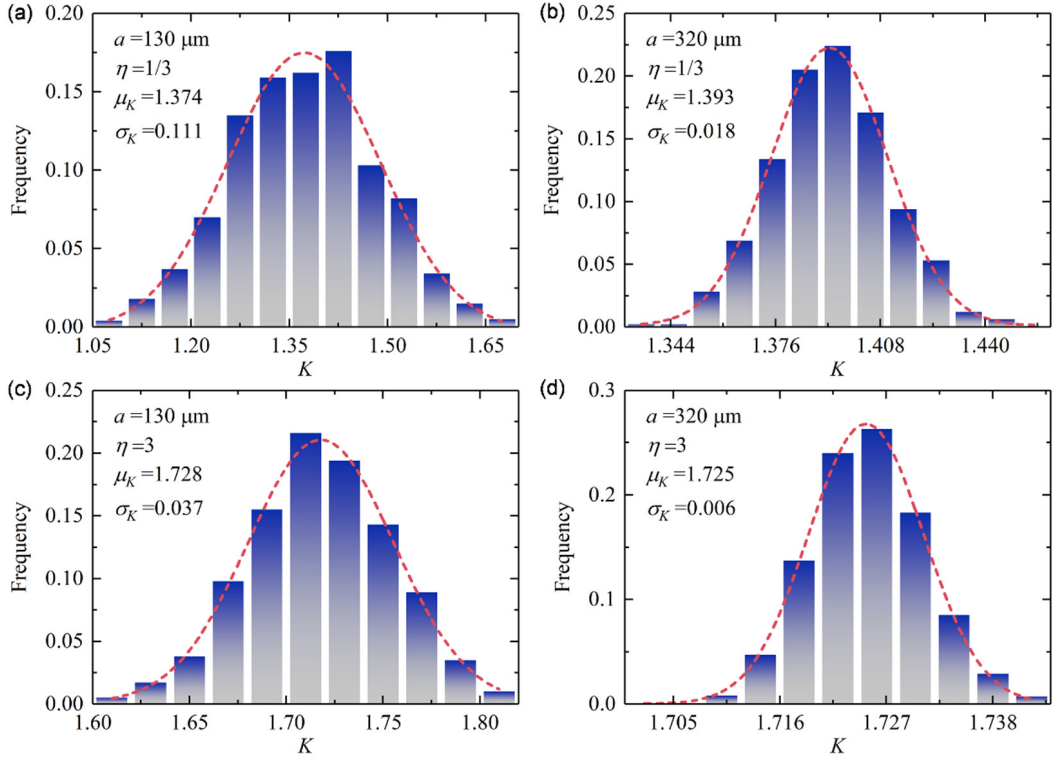

Fig. 19 Probability histograms of $K$ with different RVE edge length $a$ under different stress triaxialities $\eta$

Statistically, the mean and standard deviation of $K$ in different cases are influenced by the RVE edge length and the stress triaxiality. Therefore, by understanding the pattern of how $K$ changes with the RVE edge length $a$ and the triaxiality $\eta$, one can grasp the influence of $a$ and $\eta$ on creep damage evolution.

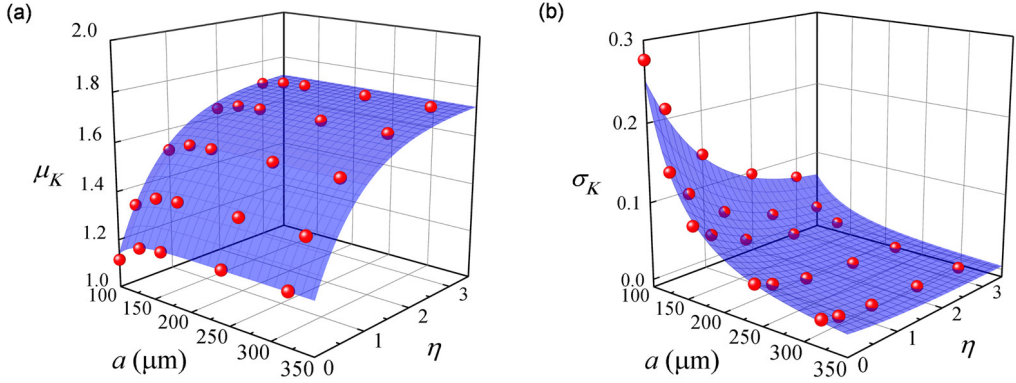

Fig. 20 The influence of RVE size and stress triaxiality on the mean $\mu_K$ and std $\sigma_K$ of creep damage rate factor $K$ (the blue surfaces represent the fitted surfaces)

The influence of $a$ and $\eta$ on the mean value and standard deviation of $K$ are shown in Fig. 20. As $\eta$ increases, there is a corresponding rise in $k$, which subsequently results in an augmentation of the mean of $K$ (see Fig. 20a). Meanwhile, the dispersion introduced by $\psi_i$ decrease with the increase of $k$, leading to a reduce of standard deviation $\sigma_K$ (see Fig. 20b). Nonlinear surface fitting was implemented for $\mu_K$ and $\sigma_K$. The fitted surfaces are shown in Fig. 20 with formulas expressed as:

$$\mu_K = 1.73 - 0.52\exp(-1.12\eta) - 6.98\exp(-0.05a - 1.12\eta) \tag{29}$$

$$\sigma_K = \left[3690.60\exp(-0.86\eta) + 1095.46\right] \cdot a^{-2.14} \tag{30}$$

According to Eqns. (13) and (28), the effective GB damage can be expressed as:

$$D_{\text{GB}}^{\text{eff}}(t) = B \cdot \sigma_1^q \cdot K \cdot t^p \tag{31}$$

in which $K$ is a stochastic value. Hence, for a given RVE edge length $a$ and stress triaxiality $\eta$, a stochastic effective GB damage can be obtained. Hence, a macroscopic creep damage can be calculated based on Eqn. (23).

The simulation of creep failure of structures can be conducted using FEM. The stochastic creep damage of RVE is updated at integration points of elements. As shown in Fig. 21, the key values are updated at the integration points. Initially, an elastic preload is applied to obtain the triaxiality of stress. Simultaneously, the volume of the integration points is read, and a state variable $\lambda$ conforming to a standard normal distribution is assigned at each integration point. Based on this, $K$ is updated, followed by updates to the damage rate and strain rate. Subsequently, the strain, stress, damage, and the new triaxiality of stress can be obtained. Updates continue in this manner until the calculation is completed. The implementation of the simulation is based on the commercial finite element software Abaqus, with the user subroutines USDFLD and GETVRM being used.

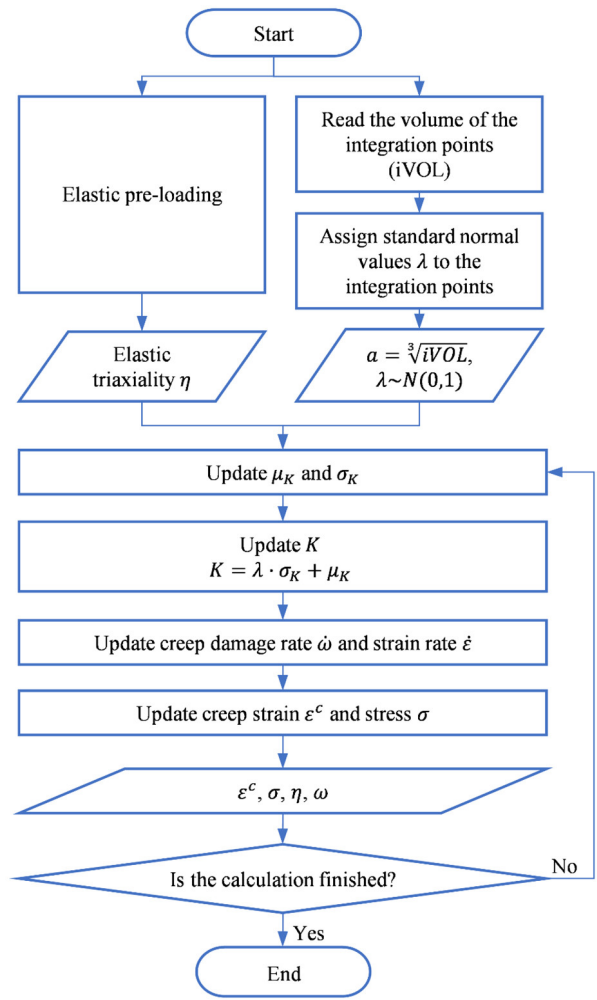

Fig. 21 Update of key variables at integration points

Hence, the inherent randomness introduced by microstructures is directly incorporated into the stochastic damage model, with the full-field simulation being fully implemented using FEM. The random distribution of GB orientations and areas is fully considered into the proposed creep damage rate factor $K$. The statistical features of $K$ are obtained from the proposed MC method.

Unlike most deterministic homogenization models, the stochastic creep damage model proposed effectively captures the influence of microstructural variations. It is noted that the stochastic characteristics of the parameter $K$ are associated with the mesh size used in the FEM. Physically, as the mesh size increases, the number of underlying GBs at the integration points of the elements increases, leading to a reduction in the stochastic nature of material behavior. This correlation between stochasticity and mesh size has also been observed in existing studies (Gorgogianni et al., 2022; Vievering and Le, 2024).

Even though the modeling process is complex, the proposed dual-scale stochastic framework is highly efficient. Its high efficiency stems from two aspects. (1) the simplification of the microstructure reconstruction: the method does not require complex iterative procedures to generate a microstructure that matches the grain size and sphericity distribution, which takes about 40 minutes for an RVE with an edge length of 320 micrometers. Instead, it only involves random sampling based on the distribution of GB characteristics. The sampling process for an RVE of the same size taking approximately 0.001 seconds. (2) The low cost of MC computation: as shown in Fig. 12, the computation time for MC ranges between 0.01 to 10 seconds, depending on the scale of the problem being solved, whereas the computation based on CPFE-CZM requires more than 20 hours. The high efficiency comes at the cost of sacrificing detailed microstructural deformation and failure behavior.

## 5.2. Results of stochastic creep failure of typical specimens

The creep deformation and failure behaviors can be simulated based on the stochastic model proposed. As the typical specimen types, the uniaxial tensile specimen and pressurized tube made of Inconel 617 is simulated and analyzed.

**5.3.1 The uniaxial tensile creep failure**

Based on the proposed stochastic creep damage model, the calculation of the creep deformation and failure behavior can be conducted. The FEM modeling and mesh convergence verification are presented in Appendix E. The curve of creep strain versus creep time is obtained from FEM calculation based on Eqn. (31) and shown in Fig. 22. Meanwhile, the creep failure under the uniform stress state can be predicted by the MC calculation of the RVE. The uniaxial creep curve obtained from MC calculation is also shown in Fig. 22. It is found that the FEM results have great agreement with the MC calculation and the experimental curve.

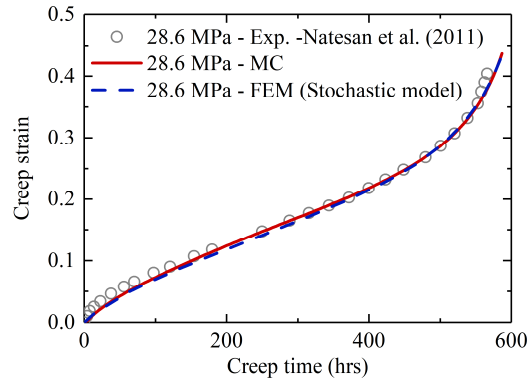

Fig. 22 The creep strain versus creep time for uniaxial tensile creep with $\sigma_e = 28.6$ MPa

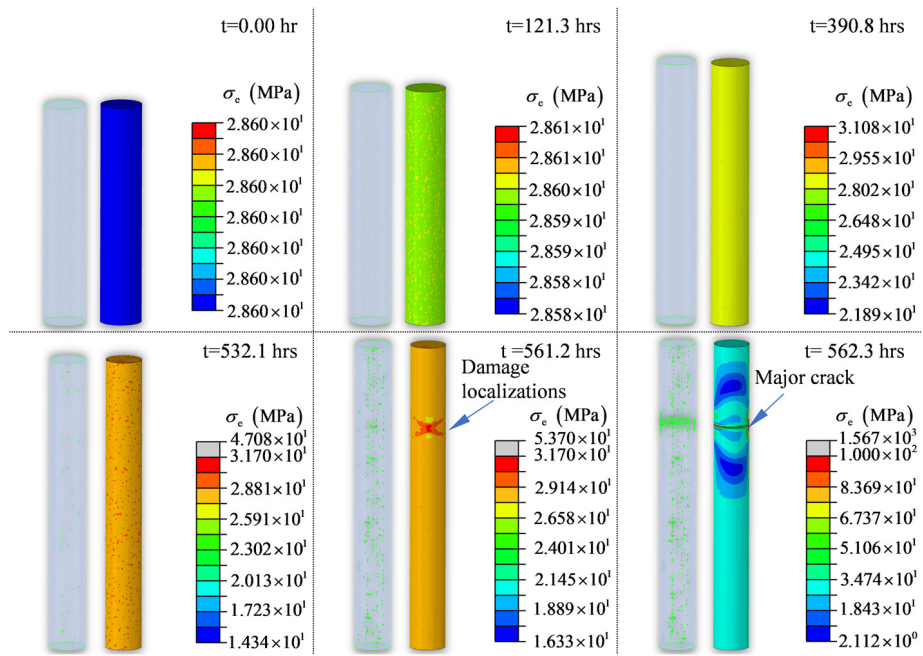

Fig. 23 The evolution of random creep damage and failure under uniaxial tension: transparent view and distribution of the von Mises equivalent stress at different creep time $t$

The random failure process and damage evaluation under uniaxial tension is shown in Fig. 23. Before the onset of creep, a uniform elastic loading was applied. As the creep progresses ($t = 121.3$ hours), the stress at each material point becomes non-uniform, due to differences of microstructures at the integration points. When the creep time reaches 390.8 hours, the non-uniformity of stress becomes more pronounced, with the maximum von Mises equivalent stress at 31.08 MPa and the minimum at 21.89 MPa, deviating from the applied 28.6 MPa. At 532.1 hours, deletion of elements occurred due to reaching the critical damage value, leading to the random initiation of microcracks in the specimen. Subsequently (561.2 hours), as creep damage developed further, localization of damage occurred, leading to significant cracks larger than others. Meanwhile, the stress concentrations became more pronounced. Ultimately, one of the significant cracks evolved into a major crack, traversing the specimen and causing it to fracture.

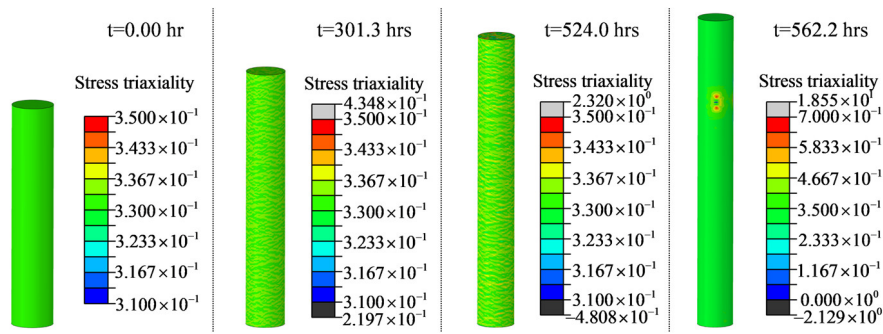

Fig. 24 The distribution of stress triaxiality under creep tension

Furthermore, the stress within the specimen is not strictly uniform due to the influence of micro structure, even though it is subjected to uniaxial tension. The corresponding stress triaxiality is shown in Fig. 24. Under a strictly uniform uniaxial tensile stress state, the stress triaxiality is equal to $1/3$. When creep happens, the stress triaxiality of the specimen fluctuates around $1/3$ due to the influence of the microstructure. The greater the creep deformation, the more pronounced the non-uniformity of the stress triaxiality. Until a major crack appears, the triaxiality significantly changes near the crack. It is noted that the MC approach only assess the creep behavior and failure under uniform stress state. However, the FEM can simulate creep failure under complex non-uniform stress state. Hence, there is a difference between the MC predicted and the FEM simulated curves.

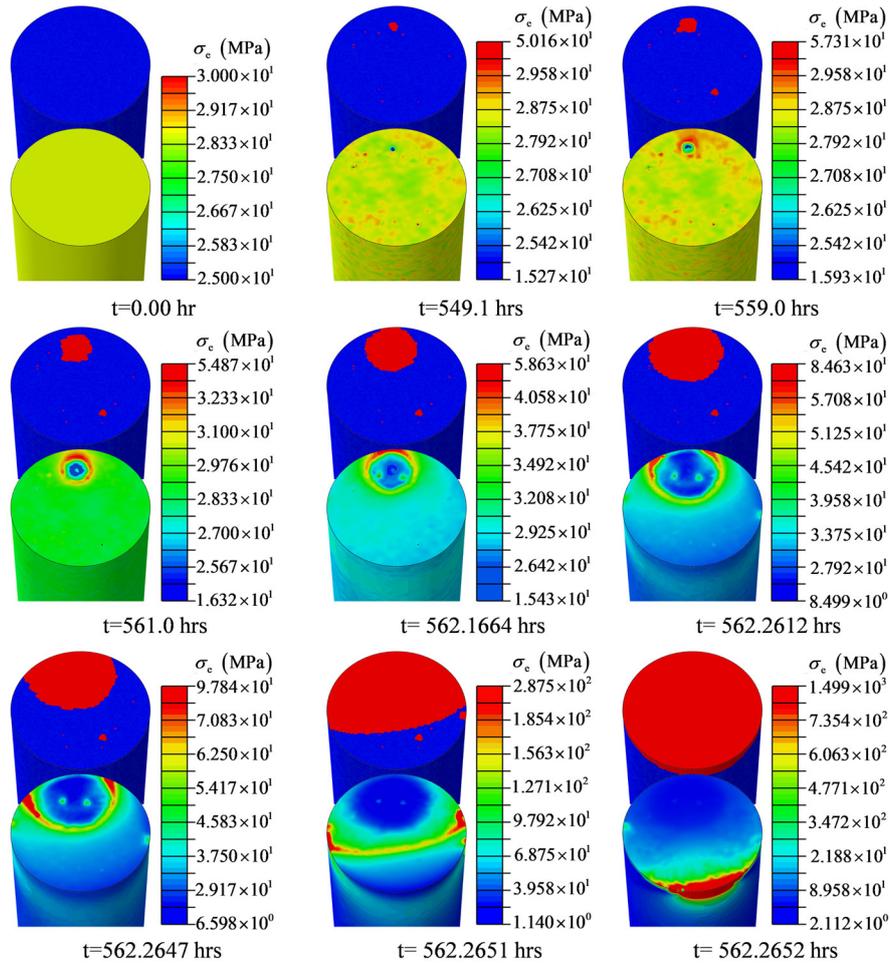

Fig. 25 The major crack propagation and von Mises equivalent stress distribution at the rupture section

The propagation process of major crack and the distribution of von Mises equivalent stress are shown in Fig. 25. The red part in the sub-figure means the cracked surface. In this study, crack initiates randomly. It can be observed that cracks initiates within the specimen. Significant stress concentration can be observed at the crack tips. As the creep time increases, the crack propagates to the surface of specimen and traverses the entire specimen. It should be noted that this study does not take into account factors such as the surface condition of the specimen and oxidation, which may affect the initiation of cracks.

### 5.3.2 The creep failure of pressurized tube

The pressurized tube creep test is used to study the creep failure of material under the biaxial stress state (Tung et al., 2014; Wright and wright, 2013). The creep deformation and failure behavior of pressurized tube can be simulated through FEM based on Eqn. (31). The FEM modeling and the verification of mesh convergence are shown in Appendix F.

The equivalent strain versus creep time is shown in Fig. 26a. It is found that the FEM results agree well with the experimental curve. However, there is an apparent difference between the MC calculated results and the experimental curves. The MC approach can only assess the failure under uniform stress states. However, the actual stress state of pressurized tube is complexly multiaxial as shown in Fig. 27. There is a significant stress gradient along the thickness direction of the tube. Only on the outer surface of the tube does the stress triaxiality remain around $\sqrt{3}/3$. FEM can reflect the non-uniform stress state in the specimen, achieving better results than MC calculation. This indicates that the actual load and structural geometry affect the uniformity of stress in the experiment. The creep rupture time of the pressurized tube under different effective stresses is shown in Fig. 26b. Here, the rupture of pressurized tube is defined as the situation that the crack propagated through the thickness of the tube, causing leakage. Similarly, the FEM calculation results agree better with the experimental results. Since the dominant stress state for the tube specimen is biaxial, the creep life predicted by MC method is also relatively accurate.

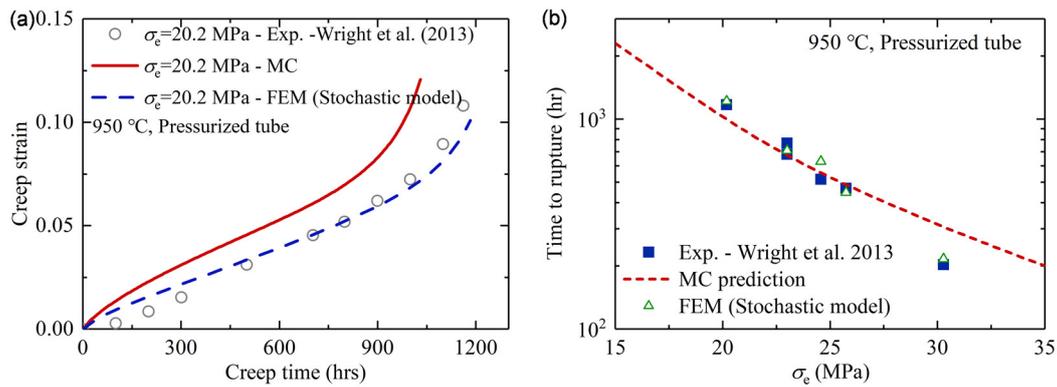

Fig. 26 (a) The equivalent strain versus creep time (b) The experimental and simulated creep rupture time

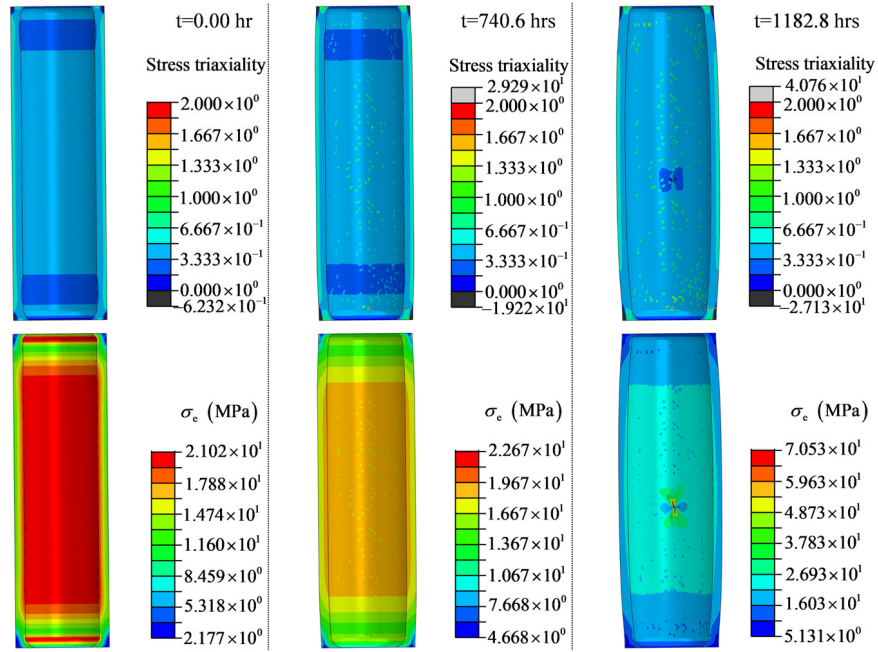

Fig. 27 The distribution of stress triaxiality and von Mises equivalent stress on the cross-section of pressurized tube

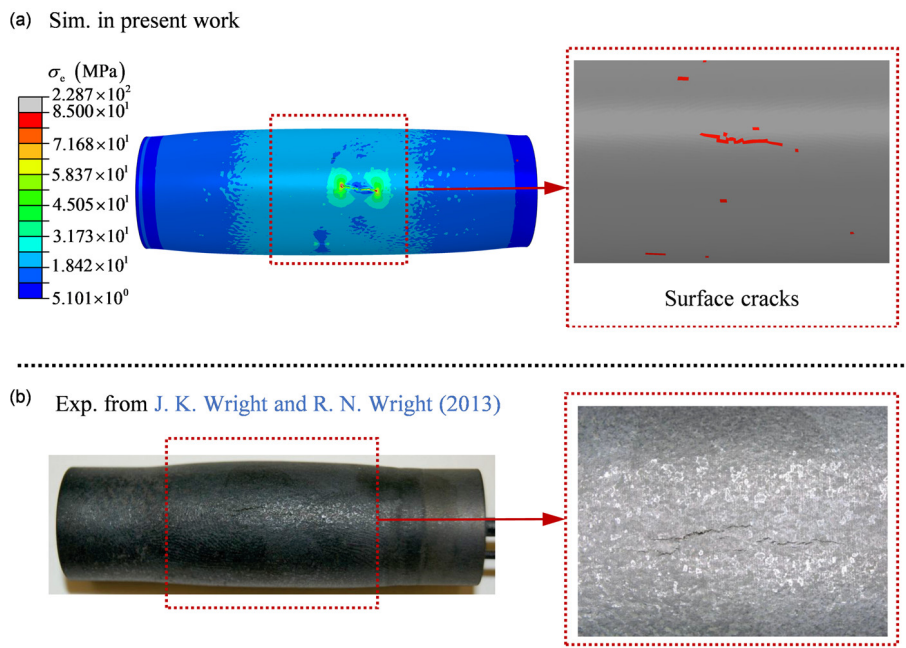

Fig. 28 The failure morphology of the pressurized tube in (a) simulation and (b) experiment

The failure morphologies of the pressurized tube in simulation and experiment are shown in Fig. 28. As shown in Fig. 28a, the failed tube surface is covered with creep cracks, and there is a major crack surrounded by a typical butterfly-shaped equivalent stress zone. As shown in Fig. 28b, the failure mode in the simulation is consistent with that observed in the experiment (Wright and wright, 2013). The propagation of the major crack occurs axially along the tube, which is consistent with its stress state. In other words, the crack plane is perpendicular to the direction of the maximum principal stress, exhibiting characteristics of brittle fracture. Due to the stochastic nature of the proposed damage model, the simulation effectively captures the stochastic initiation and propagation of surface creep cracks.

Furthermore, as shown in Fig. 29, The evaluation of random creep damage and failure of pressurized tube is simulated. The three-dimensional distributions of stochastic stress and creep damage are obtained. The green dots represent the micro-cracks consisting of failed elements. The micro-cracks primarily occur in the middle of the pressurized tube. In fact, in the creep failure calculations conducted with different randomly distributed characteristics, although the crack initiation locations are different, the internal pressure tubes exhibit similar failure patterns. In pressurized tubes with different random distribution characteristics, cracks always initiate in the middle of the tube and propagate along the axial direction of the specimen.

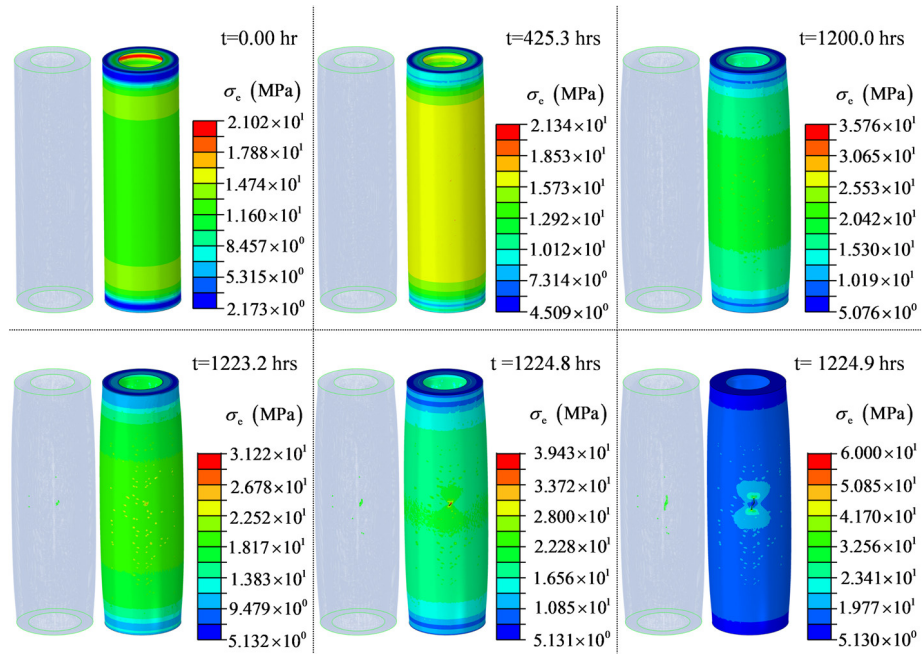

Fig. 29 The evaluation of random creep damage and failure of pressurized tube: transparent view and distribution of the von Mises equivalent stress

## 6. Conclusions

In this paper, a dual-scale stochastic analysis framework is established to study the influence of microstructure on the random creep damage and failure for superalloy Inconel 617. Firstly, the failure mechanism of GBs is revealed through CPFE-CZM calculation. Subsequently, a novel MC approach is proposed to characterize the relationship between the distribution of GB features and creep failure. Finally, a dual-scale stochastic creep damage model is proposed and used to numerically investigate the random creep damage and failure of structures. Specific conclusions are drawn as follows:

1. At the microscale, the distribution characteristics of GBs and the mechanisms of creep failure are revealed for Inconel 617. As the main microstructural features controlling creep, the distribution characteristics of the GB orientation and areas are statistically presented based on the reconstructed three-dimensional microstructure. The creep failure mechanisms of GBs are investigated through CPFE-CZM simulations. The simulation results indicate that GB damage and failure primarily occur at GBs perpendicular to the direction of maximum principal stress. The evolution of damage is highly associated with the effective normal stress on the GB surfaces.

2. A high-efficient MC approach is proposed as the bridge from microscopic mechanism to macroscopic creep behavior. When the number of GBs used for MC calculation is sufficiently large, the properties of the RVE can be considered to converge to the properties of the material. Based on the MC approach, equivalent creep strain curves and creep lifetimes under uniaxial, biaxial, and multiaxial uniform stress state are obtained. The MC calculation results for uniaxial and biaxial conditions agree well with experimental results. The calculation results for multiaxial conditions also correspond well with well-known multiaxial failure models.

3. To facilitate the analysis of creep failure and life prediction for structures containing different stress states at the macroscopic scale, a dual-scale stochastic creep damage model is established. The model is numerically implemented through FEM. It is noted that the dispersion of material properties among integration points is related to the volume and stress state at each integration point. The larger the volume corresponding to an integration point is, the smaller the variability in creep properties among the material points will be. Similarly, the higher the stress triaxiality is, the smaller the variability in creep properties will be. The inherent randomness introduced by microstructural variability has been integrated into the FEM calculation process.

4. Full-field calculations and analysis on uniaxial creep tensile specimens and pressurized tubes are efficiently conducted. The simulated results agree well with the experimental data. It is found that the stress state within the specimens is non-uniform due to microstructural variability. This also leads to FEM calculation results being generally more accurate than MC predictions in structural creep failure analysis, especially for pressurized tubes. Meanwhile, the FEM calculation also captures the random initiation and propagation of creep cracks within the specimen. For uniaxial creep tensile specimens, cracks initiate internally and propagate to the surface, eventually traversing the entire specimen. For pressurized tubes, the major crack grows along the axial direction of the tube.

Overall, this study thoroughly investigates the influence of microstructure on the creep deformation and failure behavior of Inconel 617. The characteristic distribution of GBs is considered and incorporated in the assessment of creep lifetime through a MC approach. The further developed dual-scale stochastic creep damage model enables computational analysis and failure prediction of structures at the macroscopic scale. The dual-scale stochastic analysis framework proposed in this paper provides a foundation for the investigation of creep fracture mechanisms and integrity analysis of superalloys and structures at high temperatures.

## Acknowledgments

The study was supported by the Major Project of the National Natural Science Foundation of China (12090033) and partly supported by the Key Project of the National Natural Science Foundation of China (12332005). The first author would like to thank Dr. Jia Sun from China National Petroleum Corporation for the valuable discussions.

## Appendix A The CPFE-CZM model

The CPFE framework proposed by Marin and Dawson (1998) is adopted to describe the plastic deformation of each single crystal. In this framework, the plastic deformation is assumed to be only induced by crystallographic slip. Hence, the deformation gradient $\mathbf{F}$ is decomposed into the purely plastic deformation gradient $\mathbf{F}^{p}$ induced by slide, the elastic lattice rotation $\mathbf{R}^{e}$ and the symmetric left elastic stretch tensor $\mathbf{V}^{e}$ in sequence:

$$\mathbf{F} = \mathbf{V}^{e} \cdot \mathbf{R}^{e} \cdot \mathbf{F}^{p} \tag{A.1}$$

As shown in Fig. 2a, $\mathbb{C}_0$ and $\mathbb{C}$ are the initial and current configurations, respectively. And there are two intermediate configurations introduced by Eqn. (A.1), i.e., $\bar{\mathbb{C}}$ defined by $\mathbf{F}^{p}$ and $\tilde{\mathbb{C}}$ obtained by elastically unloading from current configuration $\mathbb{C}$. The variables and equations of the CPFE formulations will be expressed and established in the relaxed configuration $\tilde{\mathbb{C}}$. The unit vectors along the $\alpha$-slip direction are denoted as $\mathbf{s}^{\alpha}$, $\bar{\mathbf{s}}^{\alpha}$ and $\tilde{\mathbf{s}}^{\alpha}$, the unit vectors normal to the $\alpha$-slip plane are denoted as $\mathbf{m}^{\alpha}$, $\bar{\mathbf{m}}^{\alpha}$ and $\tilde{\mathbf{m}}^{\alpha}$. As shown in Fig. 2a, the relation between $(\bar{\mathbf{s}}^{\alpha}, \bar{\mathbf{m}}^{\alpha})$ and $(\tilde{\mathbf{s}}^{\alpha}, \tilde{\mathbf{m}}^{\alpha})$ is written as:

$$\begin{aligned} \tilde{\mathbf{s}}^{\alpha} &= \mathbf{R}^{e} \cdot \bar{\mathbf{s}}^{\alpha} \\ \tilde{\mathbf{m}}^{\alpha} &= \mathbf{R}^{e} \cdot \bar{\mathbf{m}}^{\alpha} \end{aligned} \tag{A.2}$$

Hence, the Schmid tensors in configurations $\bar{\mathbb{C}}$ and $\tilde{\mathbb{C}}$ are denoted as $\bar{\mathbf{s}}^{\alpha} \otimes \bar{\mathbf{m}}^{\alpha}$ and $\tilde{\mathbf{s}}^{\alpha} \otimes \tilde{\mathbf{m}}^{\alpha}$, respectively. The velocity gradient $\tilde{\mathbf{L}}^{\mathrm{p}}$ in configuration $\tilde{\mathbb{C}}$ is expressed as:

$$\tilde{\mathbf{L}}^{\mathrm{p}} = \dot{\mathbf{R}}^{\mathrm{e}} \cdot \left(\mathbf{R}^{\mathrm{e}}\right)^{\mathrm{T}} + \mathbf{R}^{\mathrm{e}} \cdot \dot{\mathbf{F}}^{\mathrm{p}} \cdot \left(\mathbf{F}^{\mathrm{p}}\right)^{-1} \cdot \left(\mathbf{R}^{\mathrm{e}}\right)^{\mathrm{T}} \tag{A.3}$$

And it can be decomposed into the symmetric deformation rate $\tilde{\mathbf{D}}^{\mathrm{p}}$ and skew-symmetric spin $\tilde{\mathbf{W}}^{\mathrm{p}}$:

$$\tilde{\mathbf{L}}^{\mathrm{p}} = \tilde{\mathbf{D}}^{\mathrm{p}} + \tilde{\mathbf{W}}^{\mathrm{p}} \tag{A.4}$$

It should be noted that both the glide and climb should be considered for Inconel 617 under high temperature. The combination of glide and climb was conducted by (Phan et al., 2017), i.e.,

$$\tilde{\mathbf{D}}^{\mathrm{p}} = \tilde{\mathbf{D}}_{\mathrm{g}}^{\mathrm{p}} + \tilde{\mathbf{D}}_{\mathrm{c}}^{\mathrm{p}} \tag{A.5}$$

$$\tilde{\mathbf{W}}^{\mathrm{p}} = \tilde{\mathbf{W}}_{\mathrm{g}}^{\mathrm{p}} + \tilde{\mathbf{W}}_{\mathrm{c}}^{\mathrm{p}} \tag{A.6}$$

According to Eqns. (A.3) and (A.4), the glide part can be expressed as:

$$\tilde{\mathbf{D}}_{\mathrm{g}}^{\mathrm{p}} = \sum_{\alpha=1}^{n} \dot{\gamma}_{\mathrm{g}}^{\alpha} \left(\tilde{\mathbf{s}}^{\alpha} \otimes \tilde{\mathbf{m}}^{\alpha}\right)_{\mathrm{S}} \tag{A.7}$$

$$\tilde{\mathbf{W}}_{\mathrm{g}}^{\mathrm{p}} = \dot{\mathbf{R}}^{\mathrm{e}} \cdot \left(\mathbf{R}^{\mathrm{e}}\right)^{\mathrm{T}} + \sum_{\alpha=1}^{n} \dot{\gamma}_{\mathrm{g}}^{\alpha} \left(\tilde{\mathbf{s}}^{\alpha} \otimes \tilde{\mathbf{m}}^{\alpha}\right)_{\mathrm{A}} \tag{A.8}$$

in which $(\cdot)_{\mathrm{S}}$ and $(\cdot)_{\mathrm{A}}$ represent the symmetric and skew-symmetric parts of the tensor. The climb part can be expressed as (Lebensohn et al., 2010; Phan et al., 2017):

$$\tilde{\mathbf{D}}_{\mathrm{c}}^{\mathrm{p}} = \sum_{\alpha=1}^{n} \dot{\gamma}_{\mathrm{c}}^{\alpha} \left(\tilde{\mathbf{K}}^{\alpha}\right)_{\mathrm{S}} \tag{A.9}$$

$$\tilde{\mathbf{W}}_{\mathrm{c}}^{\mathrm{p}} = \sum_{\alpha=1}^{n} \dot{\gamma}_{\mathrm{c}}^{\alpha} \left(\tilde{\mathbf{K}}^{\alpha}\right)_{\mathrm{A}} \tag{A.10}$$

in which

$$\tilde{\mathbf{K}}^{\alpha} = \tilde{\mathbf{m}}^{\alpha} \otimes \tilde{\boldsymbol{\chi}}^{\alpha} \tag{A.11}$$

$$\tilde{\boldsymbol{\chi}}^{\alpha} = \tilde{\mathbf{n}}^{\alpha} \times \tilde{\mathbf{t}}^{\alpha} \tag{A.12}$$

where $\tilde{\mathbf{t}}^{\alpha}$ is the tangent to the dislocation line.

It is widely accepted that the elastic strain is a relatively tiny variable compared with the in-elastic strain. With the small elastic strain assumption, $\mathbf{V}^e$ is expressed as:

$$\mathbf{V}^e = \mathbf{I} + \boldsymbol{\epsilon}^e, \quad \|\boldsymbol{\epsilon}^e\| \ll 1 \tag{A.13}$$

Hence, it is obtained:

$$\dot{\mathbf{V}}^e = \dot{\boldsymbol{\epsilon}}^e, \quad (\mathbf{V}^e)^{-1} = \mathbf{I} - \boldsymbol{\epsilon}^e \tag{A.14}$$

The Kirchhoff stress $\tau$ is adopted in the elastic constitutive equations:

$$\boldsymbol{\tau} = \tilde{\mathbb{L}} : \boldsymbol{\epsilon}^e \tag{A.15}$$

where $\tilde{\mathbb{L}}$ is the fourth order elasticity tensor of crystal and the Kirchhoff stress $\boldsymbol{\tau} = \det(\mathbf{I} + \boldsymbol{\epsilon}^e)\boldsymbol{\sigma}$.

As the key part of the CPFE model, the flow rule proposed by Busso et al. (2000) was widely used to describe the dislocation slip shearing deformation of Ni-based superalloy at high temperature (Guo et al., 2020; Guo et al., 2022; Phan et al., 2017; Zhang and Oskay, 2016):

$$\dot{\gamma}_g^\alpha = \dot{\gamma}_0 \exp\left\{-\frac{F_0}{k\theta}\left\langle 1 - \left\langle \frac{|\tau^\alpha - B^\alpha| - S^\alpha \mu/\mu_0}{\hat{\tau}_0 \mu/\mu_0}\right\rangle^{p_g}\right\rangle^{q_g}\right\} \operatorname{sgn}(\tau^\alpha - B^\alpha) \tag{A.16}$$

in which $\tau^\alpha$ is the resolved shear stress:

$$\tau^\alpha = \boldsymbol{\tau} : (\tilde{\mathbf{s}}^\alpha \otimes \tilde{\mathbf{m}}^\alpha)_S \tag{A.17}$$

And $\dot{\gamma}_0$, $k$, $\theta$ and $F_0$ are the reference shear strain rate, Boltzmann constant, Kelvin temperature and the activation energy, respectively. $\hat{\tau}_0$ is the threshold stress over which the dislocation can be mobilized without thermal activation. $\mu$ is the shear moduli at current temperature, and $\mu_0$ is the shear moduli at 0 K. $S^\alpha$ is the slip resistance to dislocation motion proposed by Zhang and Oskay (2016) for Inconel 617 at high temperature, and its evolution obeys:

$$\dot{S}^\alpha = \left[h_S - d_D(S^\alpha - \bar{S}^\alpha)\right]|\dot{\gamma}^\alpha| - h_2(S^\alpha - S_0^\alpha)H\left(\dot{\gamma}_{\mathrm{th}} - \sum_{\alpha=1}^n |\dot{\gamma}^\alpha|\right) \tag{A.18}$$

in which $h_S$ and $d_D$ represent the hardening and dynamic recovery parameters. $\bar{S}^\alpha$

is the steady flow strength parameter of the $\alpha$-slip system. $h_2$ is the static recovery parameter. $S_0^\alpha$ and $\dot{\gamma}_{th}$ represent the initial slip resistance and the threshold rate for static recovery, respectively. $H(\bullet)$ represents the Heaviside function.

The evolution function of back stress $B^\alpha$ in Eqn. (A.16) proposed by Lin et al. (2010) is expressed as:

$$\dot{B}^\alpha = \left[ h_B - D^\alpha B^\alpha \operatorname{sgn}(\dot{\gamma}^\alpha) \right] \dot{\gamma}^\alpha \tag{A.19}$$

where $h_B$ and $D^\alpha$ are the hardening and dynamic recovery parameters for back stress with $D^\alpha$ expressed as:

$$D^\alpha = \frac{h_B \mu_0}{S^\alpha} \left( \frac{\mu_0'}{f_{co}} - \mu \right)^{-1} \tag{A.20}$$

in which $\mu_0'$ is the local slip shear modulus at 0 K. $f_{co}$ is a coupling parameter for the internal slip variables.

Meanwhile, dislocation climb, which is important in describing the secondary creep is incorporated into the CPFE framework. Dislocation climb happens in the so-called climb system, just like dislocation glide happens in the slip system. Dislocation climb contributes to the plastic flow through the climb flow rule:

$$\dot{\gamma}_c^\alpha = \dot{\gamma}_0 \exp\left(-\frac{F_0}{k\theta}\right) \left( \frac{|\tau_c^\alpha - B_c^\alpha|}{\hat{\tau}_{0c}} \right)^{p_c} \operatorname{sgn}(\tau_c^\alpha - B_c^\alpha) \tag{A.21}$$

in which $\hat{\tau}_{0c}$ and $p_c$ are the scalar threshold creep stress and creep exponent. And the shear stress of climb system is expressed as:

$$\tau_c^\alpha = \boldsymbol{\tau} : \left( \tilde{\mathbf{K}}^\alpha \right)_{dev} \tag{A.22}$$

And the evaluation of back stress is given as follows (Staroselsky and Cassenti, 2011):

$$\dot{B}_c^\alpha = h_c \left( \dot{\gamma}_c^\alpha B_\infty - |\dot{\gamma}_c^\alpha| B_c^\alpha \right) \tag{A.23}$$

where $B_\infty$ is a saturation value and $h_c$ is a material parameter.

The CPFE framework and model elaborated in Eqns. (A.1)~(A.23) has been successfully used and reliably verified for the monotonic creep and creep-fatigue of Inconel 617 at high temperature (Phan et al., 2017; Zhang and Oskay, 2016). Hence, the framework and model mentioned above are employed for the calculation of the creep behavior of Inconel 617 at 950 °C. The parameters of CPFE glide and climb models are given in Table A 1.

Table A 1 The parameters of the CPFE model of Inconel 617 at 950°C (Zhang and Oskay (2016))

| Parameter | Value | Parameter | Value |
|---|---|---|---|
| $C_{11}$ (GPa) | 170.64 | $p_g$ | 0.181 |
| $C_{12}$ (GPa) | 108.39 | $q_g$ | 1.633 |
| $C_{44}$ (GPa) | 77.82 | $h_S$ (MPa) | 397.73 |
| $\dot{\gamma}_0$ (s$^{-1}$) | 2.288×10$^{-3}$ | $d_D$ (MPa) | 5073.62 |
| $F_0$ (J) | 5.148×10$^{-19}$ | $S_0^\alpha$ (MPa) | 143.41 |
| $\mu_0$ (GPa) | 265.33 | $\bar{S}^\alpha$ (MPa) | 18.03 |
| $\mu$ (GPa) | 77.82 | $h_B$ (MPa) | 104.31 |
| $\mu_0'$ | 31.13 | $h_2$ (MPa) | 0.015 |
| $\hat{\tau}_0$ (MPa) | 268.2 | $\dot{\gamma}_{th}$ (s$^{-1}$) | 1.0×10$^{-6}$ |
| $f_{co}$ | 0.36 | $\psi_c$ (°) | $\pi/4$ |
| $p_c$ | 3 | $\hat{\tau}_{0c}$ (Pa) | 7750 |
| $h_c$ | 32 | $B_\infty$ (MPa) | 4.7 |

The cohesive zone model has been widely used for simulated grain boundary damage in polycrystalline materials (Nassif et al., 2019; Pu et al., 2017). The cohesive law used in this paper was developed by Zhang and Oskay (2016) for Inconel 617, based on the original formulation by Bouvard et al. (2009). The potential function is expressed as:

$$\phi = \frac{1}{2\delta_c}(1-\omega_c)k_n\left(\|\mathbf{u}_n\|^2 + \alpha\|\mathbf{u}_t\|^2\right) \tag{A.24}$$

in which $\delta_c$ is the critical opening, $\omega_c$ is the creep damage. $k_n$ is the initial stiffness along the normal direction. $\mathbf{u}_n$ and $\mathbf{u}_t$ are the normal and tangential separation vectors, respectively. $\alpha$ denote a ratio between tangential and normal stiffnesses. According to the cohesive potential, the normal and shear tractions are obtained as:

$$\mathbf{T}_n = \begin{cases} k_n \left(1-\omega_c\right)\dfrac{\mathbf{u}_n}{\delta_c}, & \text{under tension} \\ k_c \dfrac{\mathbf{u}_n}{\delta_c}, & \text{under compression} \end{cases} \quad (A.25)$$

$$\mathbf{T}_t = \alpha k_n \left(1-\omega_c\right)\dfrac{\mathbf{u}_t}{\delta_c} \quad (A.26)$$

in which $k_c$ is a penalty parameter and taken as $10k_n$ here. The creep damage evaluation is assumed to be similar to the Rabotnov-Kachanov law:

$$\dot{\omega}_c = \dfrac{1}{\left(1-\omega_c\right)^{p_{dc}}} \left\langle \dfrac{\|\mathbf{T}\|-T_c}{C} \right\rangle^r \quad (A.27)$$

where

$$\|\mathbf{T}\| = \sqrt{\langle \mathbf{T}_n \rangle + \dfrac{1}{\alpha}\|\mathbf{T}_t\|^2} \quad (A.28)$$

Accordingly, the cohesive zone model can be implemented by a user defined element (UEL) in commercial software Abaqus. The use of cohesive element can realize the simulation of typical creep failure in GBs.

Meanwhile, the grains within the RVE must be sufficient to ensure the convergence of computational results. Since the modeling steps for microstructures of Inconel 617 and verification of convergence have already been demonstrated in the literature (Phan et al., 2017; Zhang and Oskay, 2016), this paper will only provide a brief explanation.

## Appendix B  Model calibration for Inconel 617 at 950 °C

For Inconel 617 at 950°C, the time hardening creep relation is fitted based on the experimental results with $\sigma_e = 18.5$ MPa and $\sigma_e = 28.6$ MPa (Natesan et al., 2012; Wright et al., 2014). Although the experimental data here come from two different reports, each being the first to present the relevant experimental data, it is important to emphasize that all the experimental data used in calibration are sourced from Idaho National Laboratory and tested at Argonne National Laboratory. Therefore, in subsequent studies, the uncertainties arising from differences in material initial defects, processing history, and testing environment can be neglected, assuming that the creep behavior obtained from macroscopic testing is accurate and exhibits good consistency. Moreover, since the macroscopic creep damage model used here is a classical and simplified formulation, it does not currently account for factors such as initial material damage, test stress, and temperature. In the future, more comprehensive damage models could be introduced to capture a broader range of effects (Hossain and Stewart, 2021). However, this would require extensive experimental calibration and parameter inversion, which is beyond the scope of this work and will not be discussed here. The calibration of time hardening creep behavior is shown in Fig. B 1 with $A = 2.675 \times 10^{-8}$, $m = 0.72$ and $n = 3.45$.

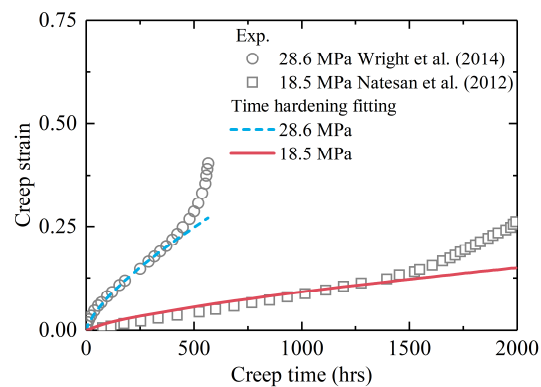

Fig. B 1 The fitting of time hardening creep relation based on uniaxial tensile experiment with $\sigma_e = 28.6,\ 18.5$ MPa

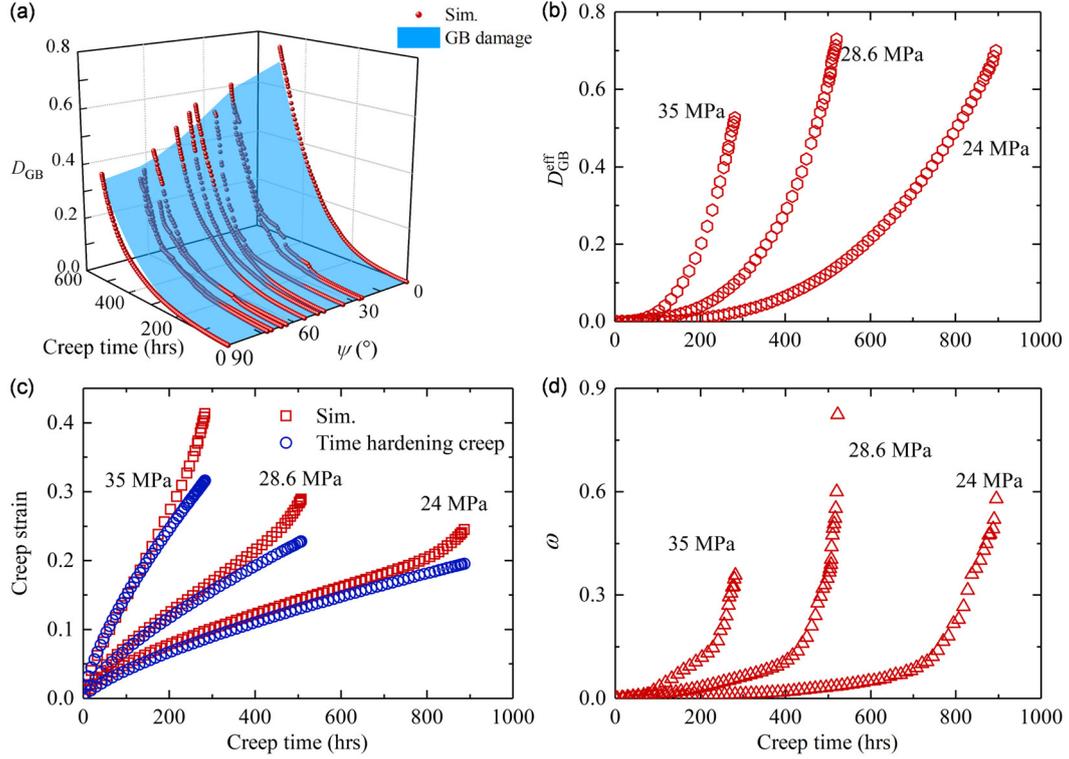

Fig. B 2 (a) The GB creep damage versus $\psi$ and creep time under uniaxial tension with $\sigma_e = 24,\ 28.6,\ 35$ MPa (b) The effective GB creep damage under uniaxial tension with $\sigma_e = 24,\ 28.6,\ 35$ MPa (c) The creep curve from CPFE-CZM calculation and time hardening curve Eqn. (14) with $\sigma_e = 24,\ 28.6,\ 35$ MPa (d) The obtained macroscopic creep damage $\omega(t)$ based on Eqn. (21) with $\sigma_e = 24,\ 28.6,\ 35$ MPa

As shown in Fig. B 2a, the GB creep damage versus angle $\psi$ is obtained from CPFE-CZM calculation of RVE, and the results are plotted in red dot line. The GB damage evaluation is assumed to be described by Eqn. (10) and plotted in blue surface diagram in Fig. B 2a. The functional form of Eqn. (10) is proven effective in describing GB creep damage. The effective GB damage $D_{GB}^{eff}(t)$ calculated through Eqn. (13) from CPFE-CZM simulation is shown in Fig. B 2b. Meanwhile, the uniaxial tensile creep curve can also be obtained from the calculation of RVE and presented with red dot line in Fig. B 2c. Accordingly, the time hardening relation Eqn. (14) is also plotted in blue dot line in Fig. B 2c. According to Eqn. (20), the macroscopic creep damage $\omega(t)$ is obtained and plotted in Fig. B 2d. Due to the difficulty in convergence when using CZM elements in calculations, it is essentially impossible to achieve macroscopic creep damage approaching 1 in CPFE-CZM simulation. With the obtained effective GB creep damage $D_{GB}^{eff}(t)$ in Fig. B 2b and macroscopic creep damage $\omega(t)$ in Fig. B 2d, a $\omega - D_{GB}^{eff}$ relation is shown in Fig. B 3 based on Eqn. (23).

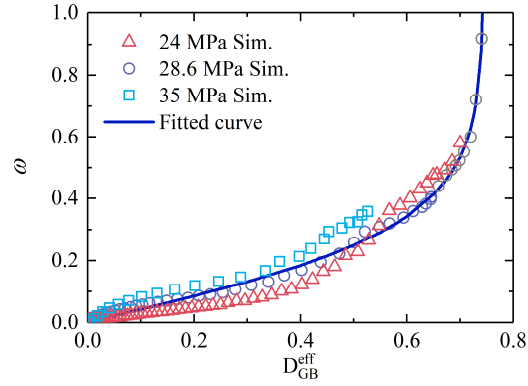

Fig. B 3 The relation between effective GB creep damage and macroscopic creep damage under different applied stress $\sigma_e = 24,\ 28.6,\ 35$ MPa

In the MC calculation process, 100 sets of GB characteristics distribution are randomly generated using the distributions shown in Fig. 7. To verify the convergence of MC calculations, cases with different RVE sizes are considered. Based on Eqns. (10) and (13), the effective GB damage of each set can be obtained and shown in Fig. B 4a. With the increasing number of GBs in each set, the dispersion range of $D_{GB}^{eff}(t)$ get narrower. For each curve of $D_{GB}^{eff}(t)$, a corresponding creep curve can be obtained according to Eqn. (20). The obtained creep curves are shown in Fig. B 4b. The averaged curves corresponding to these 100 sets of GBs are obtained and presented in Fig. B 4c. It is found that when the edge length of RVE reaches 320 μm, an expected creep curve can be obtained. The response of a macroscopic uniaxial tensile cylindrical specimen is the collective response of all the microstructures it contains. Hence, when the size of the RVE is sufficiently large, the expected creep curve is considered representative and should be consistent with the experimental curve.

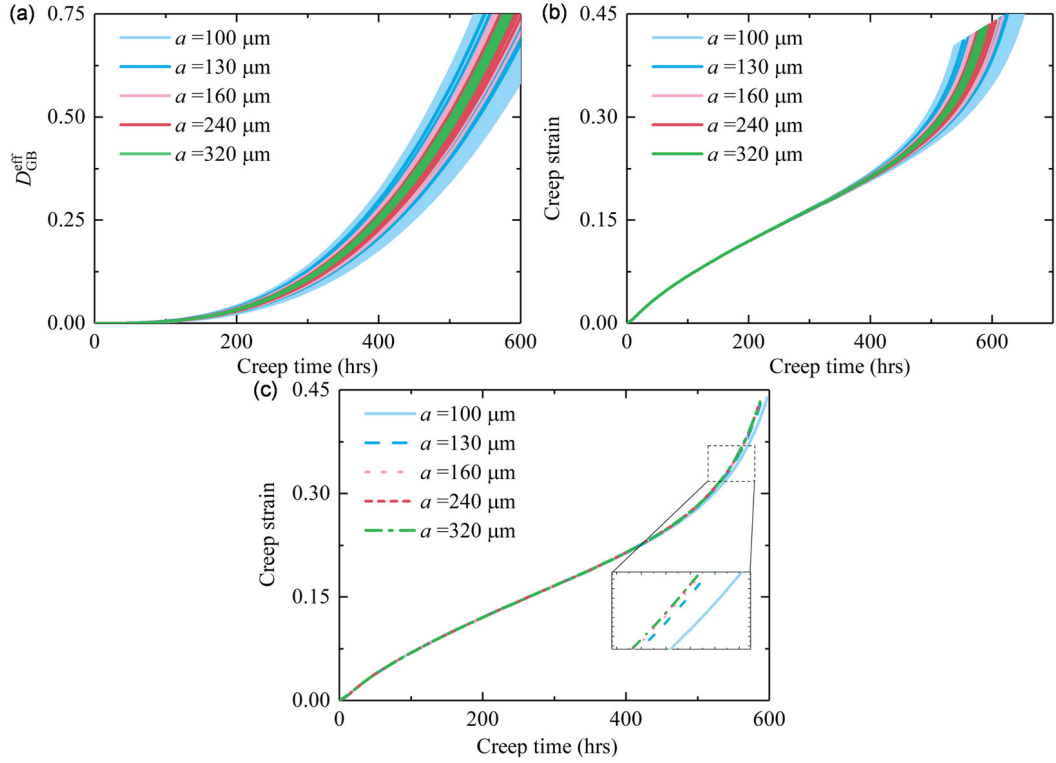

Fig. B 4 (a) The effective GB damage from Eqn. (10) and (13) in MC calculation of RVE with different side length $a$ (b) The creep curves obtained through MC calculation of RVE with different side length $a$ (c) The convergence verification of averaged creep curves in MC calculation of RVE with different side length $a$

Table B 1 The parameters of CZM and MC approach

| Parameter | Value | Parameter | Value |
| --- | --- | --- | --- |
| $\alpha$ | 0.5 | $k_n$ (Pa) | $10^{14}$ |
| $\delta_c$ (μm) | 450 | $r$ | 2.6 |
| $p_{dc}$ | 3 | $C$ (Pa) | $4.77 \times 10^{10}$ |
| $T_c$ (Pa) | $8.5 \times 10^6$ | $B$ | $2.27 \times 10^{-21}$ |
| $\beta$ | 0.63 | $p$ | 2.93 |
| $q$ | 8.43 | $A$ | $2.675 \times 10^{-8}$ |
| $m$ | 0.72 | $n$ | 3.45 |
| $a$ | 0.31 | $b$ | 1.34 |

After careful calibration, the material parameters of CZM and MC approach are obtained and given in Table B 1. The CPFE parameters are shown in Table A 1. As shown in Fig. B 5, the parameters in GB evaluation Eqn. (10) is obtained from the calibrated CPFE-CZM simulation. It can be observed that, the form of Eqn. (10) effectively describes the influence of GB orientation on GB damage under different stress levels, indicating that the semi-phenomenological Eqn. (10) is reasonable.

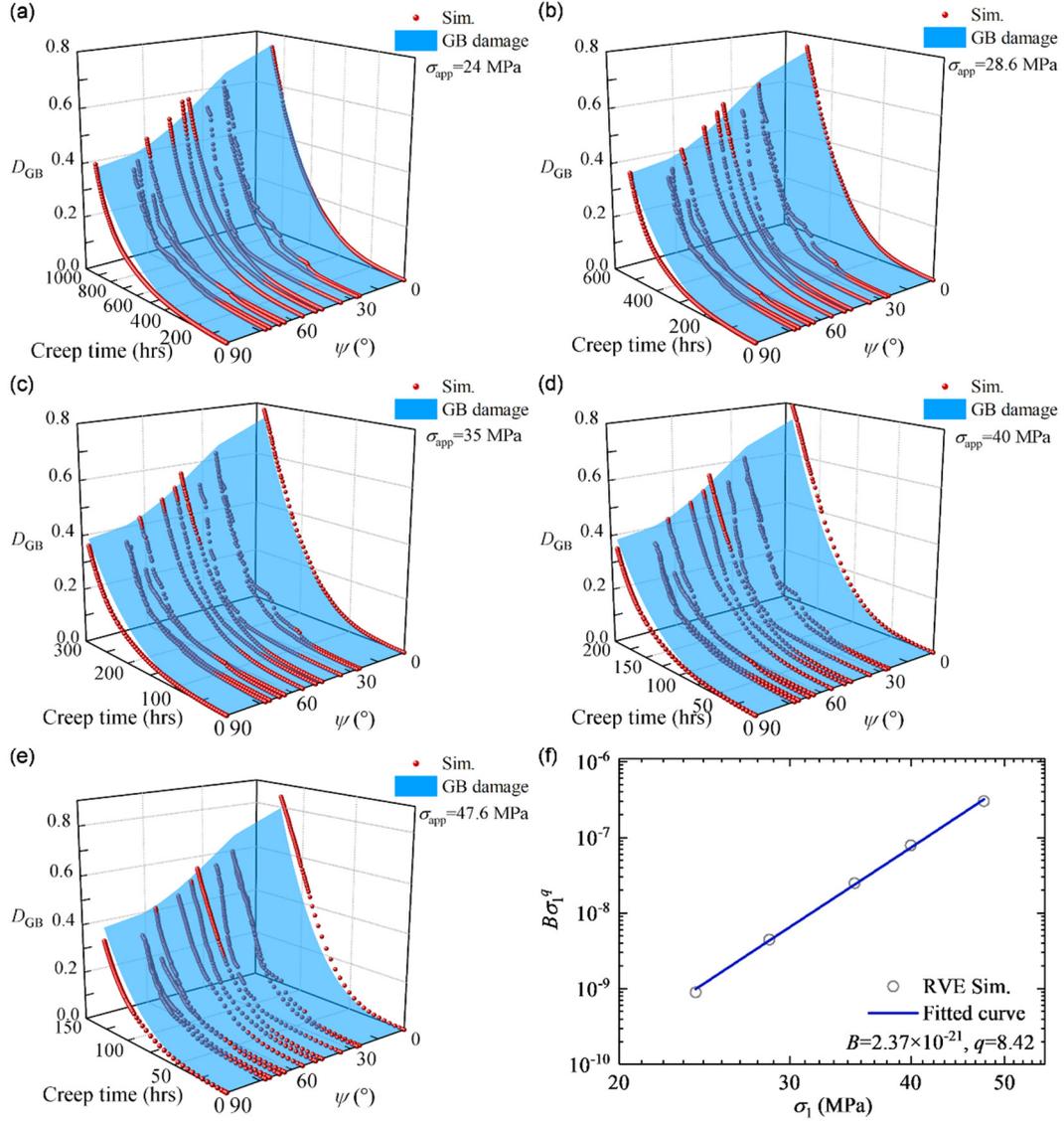

Fig. B 5 The calibration of parameters in Eqn. (10) under different applied stress, the red dot lines are from CPFE-CZM simulation, the blue surfaces are fitted according to Eqn. (10)

# Appendix C Validation of the calibrated MC model for other cases

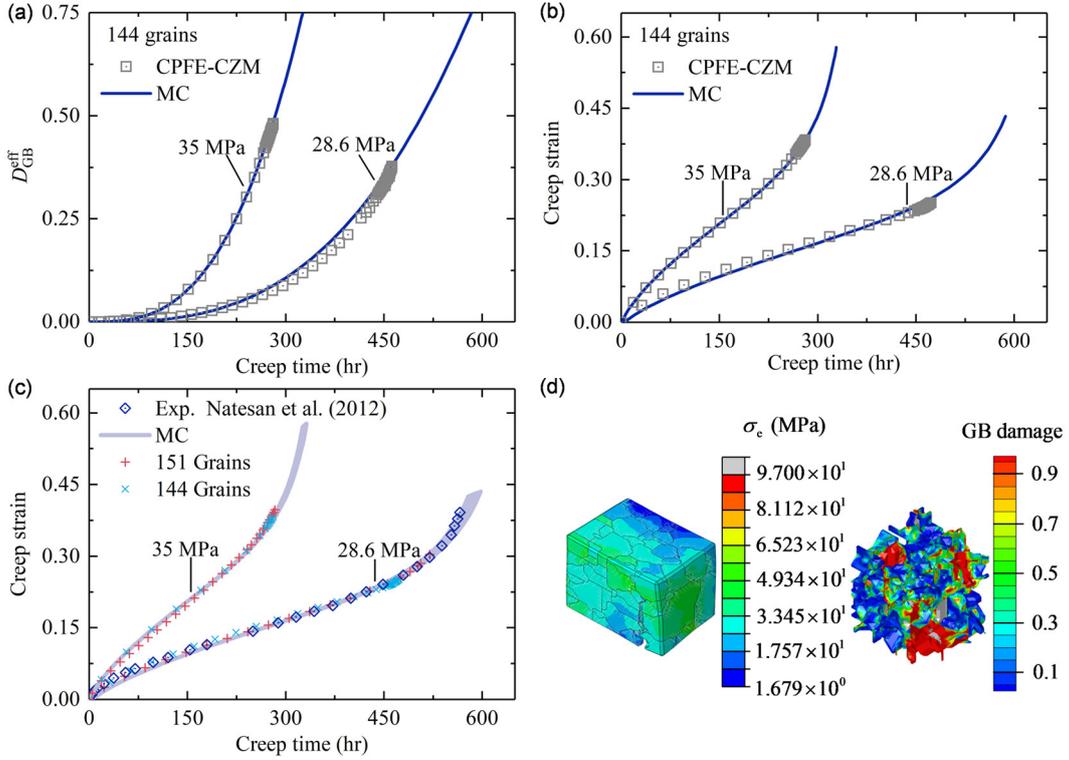

Fig. C 1 The simulation results from 144 grains model (a) effective GB damage evolution (b) creep strain curve of 144 grains model (c) comparison of MC, CPFE-CZM and experimental results (d) the stress and GB damage distribution of the RVE

The above CPFE-CZM simulation and calibration for MC approach are based on uniaxial tensile, using the same RVE configuration ($a=320$ μm, 151 grains). Here, the validity of the calibrated Eqns. (10), (20), and (23) will be verified under different RVE configurations and various stress states. As shown in Fig. C 1, the uniaxial tensile damage and deformation behavior obtained from different a RVE configuration ($a=320$ μm, 144 grains). The aim of this appendix is to validate the accuracy of the MC method for different GB configurations. Even though the number of grains is slightly different, the configurations of GBs may present significant variations. The MC results are based on the calibration in Appendix B. It is found that the CPFE-CZM results are in good agreement with the MC curves. The effective GB damage obtained from CPFE-CZM framework agrees well with the MC curves shown in Fig. C 1a, which indicates that the calibration of Eqn. (10) is reliable. As shown in Fig. C 1b, the CPFE-CZM simulated creep strain curves are in good agreement with the MC results, which indicates that the calibrations of Eqns. (20) and (23) are reliable for different RVE configurations. It is noted that due to the introduction of CZM, the computational convergence of CPFE-CZM is relatively poor, making it challenging to obtain a complete three-stage creep process. However, the computed results obtained are generally accurate. As shown in Fig. C 1c, the CPFE-CZM calculations of the different RVEs generally fall accurately within the dispersion band obtained by MC. The CPFE-CZM simulation results of grain deformation and GB damage are shown in Fig. C 1d.

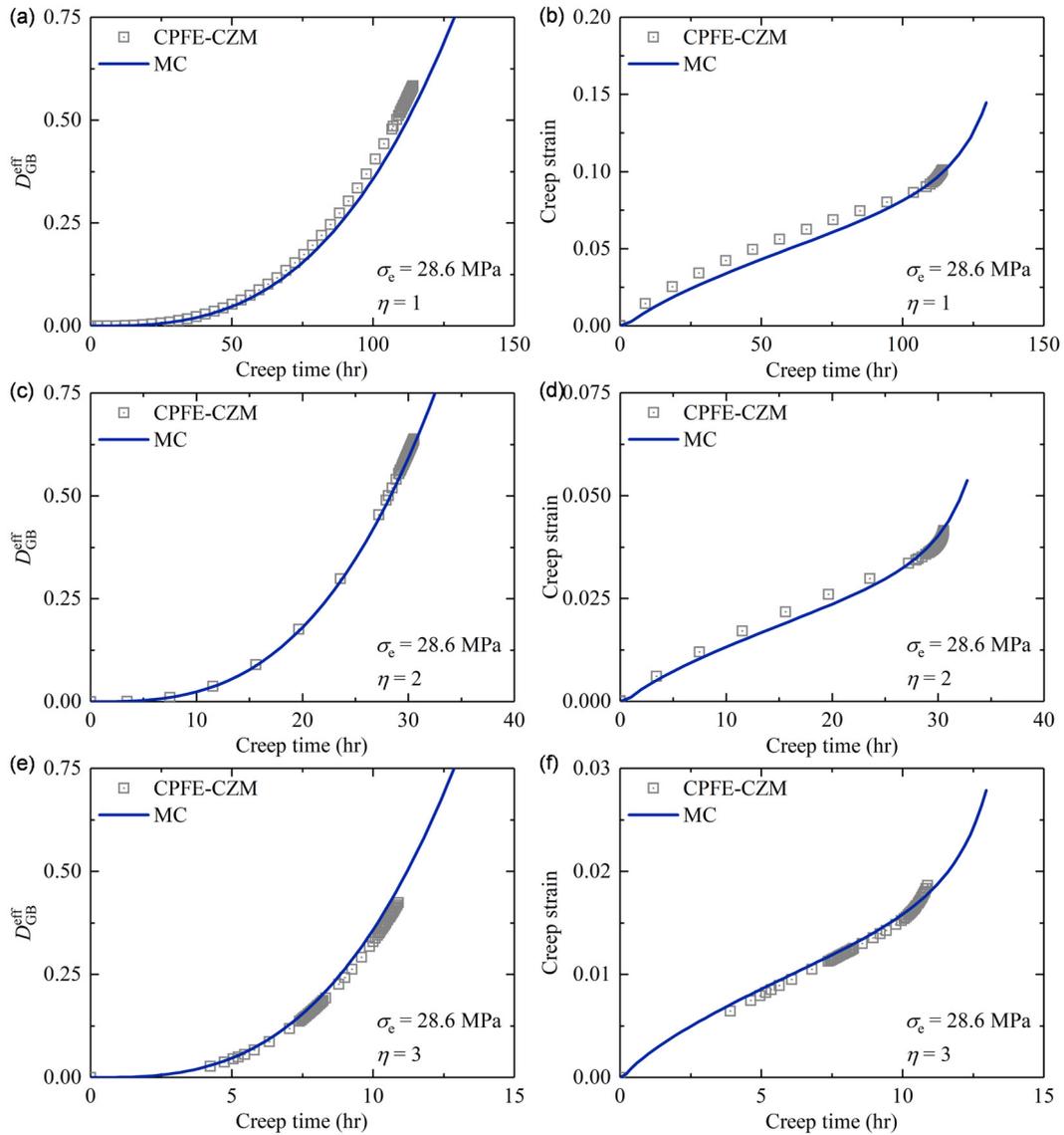

Fig. C 2 The effective GB damage evolution with $\sigma_e = 28.6$ MPa and (a) $\eta = 1$ (c) $\eta = 2$ (e)

$\eta = 3$ and the creep strain curve with $\sigma_e = 28.6$ MPa and (b) $\eta = 1$ (d) $\eta = 2$ (f) $\eta = 3$

Similarly, multiaxial CPFE-CZM simulations were also performed to validate the applicability of the calibrated model in multiaxial cases. The effective GB damage evolution and creep strain curves with $\sigma_e = 28.6$ MPa and $\eta = 1, 2, 3$ (Stress triaxiality $\eta = \sigma_m / \sigma_e$) are shown in Fig. C 2. In multiaxial cases, the CPFE-CZM simulation results show some deviations from the MC curves, but these deviations are generally acceptable. This indicates that the calibrated models Eqns. (10), (20), and (23) are also valid for multiaxial cases. The purpose of this appendix is to validate the accuracy of the MC method as a substitute for the CPFE-CZM framework, so comparisons with experimental data are limited. Due to computational cost and challenges in convergence, it is difficult to perform CPFE-CZM calculations of more cases for validations in this work.

## Appendix D The effects of RVE size and stress state on the random creep rupture behavior

1) The effect of RVE size on random creep rupture behavior

As shown in Fig. B 4, the dispersion of the creep damage and failure behavior is influenced by the size of RVEs. As a result, the creep rupture strain and time obtained from the evaluation of RVEs are related to their sizes. The distributions of creep rupture strain and time of RVEs with different sizes can be obtained through the MC calculation. The number of GB $N_{GB}$ used in MC calculation is related to the RVE edge length $a$. To obtain the relation between $N_{GB}$ and $a$, 100 different RVEs with a specific edge length $a$ are randomly reconstructed. And the corresponding values of $N_{GB}$ for these 100 RVEs are obtained. As shown in Fig. D 1, the value of $N_{GB}$ generally increases with increasing $a$. Moreover, $N_{GB}$ exhibits a certain variability for a specific value of $a$, due to the randomness of the microstructural reconstruction. However, the variability of the value of $N_{GB}$ is very limited and can be almost ignored. Therefore, for a specific RVE size with a given edge length $a$, the number of GBs used for the MC calculation is taken as the integer closest to the mean value of $N_{GB}$ shown in Fig. D 1.

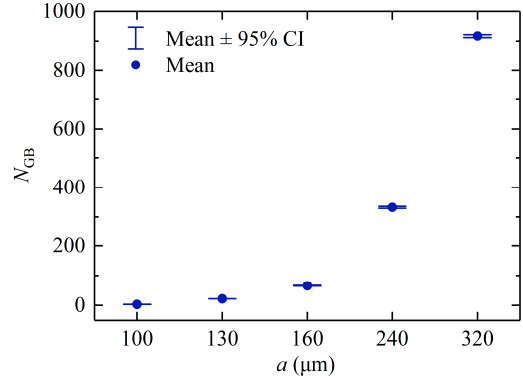

Fig. D 1 The relation between the number of GB $N_{GB}$ in RVEs with specific edge length $a$

Table D 1 The number of GB used in MC calculation for specific RVE edge length

| $a$ (μm) | 100 | 130 | 160 | 240 | 320 |
|---|---|---|---|---|---|
| $N_{GB}$ | 13 | 35 | 74 | 323 | 919 |

The calculated results are shown in Fig. D 2. Here, the uniaxial tension cases with $\sigma_e = 20$ MPa at 950 °C are taken as examples. It is observed that as the edge length $a$ of the RVE increases, the dispersion in rupture time and fracture strain decreases. Statistically, the increase of the RVE size leads to a reduction in the standard deviation. Similar phenomenon can be found in Fig. B 4a and c. This is determined by statistical theory, and this conclusion holds true for multiaxial stress conditions as well. The dispersions of the macroscopic responses of RVEs are highly related to the RVE size. This aligns with the Law of Large Numbers and Central Limit Theorem: as the size of the RVE increases, the average behavior of the element becomes more representative of the macroscopic behavior of the material. The microstructural randomness stabilizes, eventually resulting in a more concentrated and uniform macroscopic response.

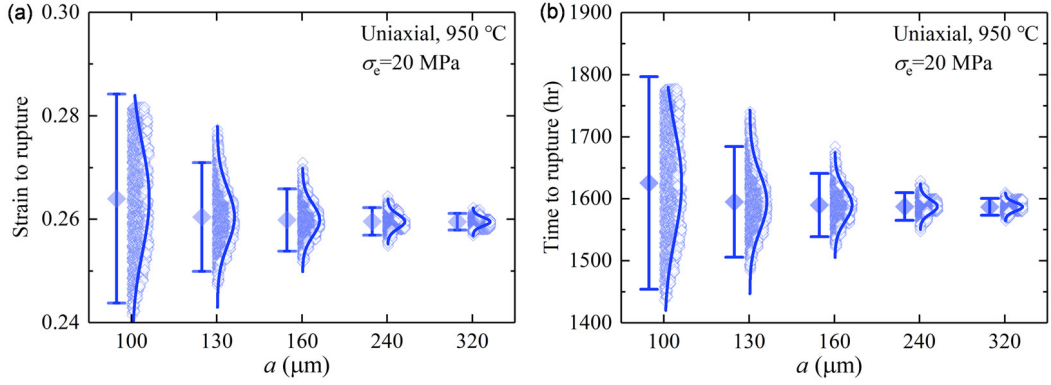

Fig. D 2 The effect of RVE size $a$ on the randomness of creep rupture time and strain (The diamond symbol represents the mean; the upper and lower lines represent the 95% confidence interval.)

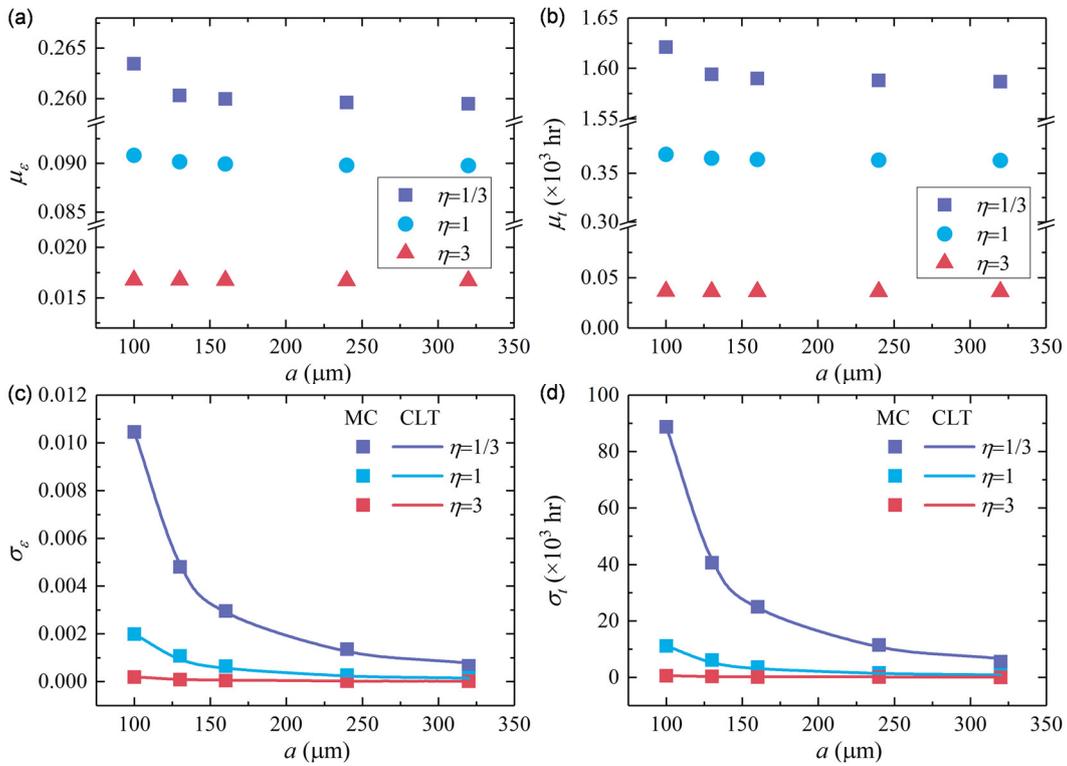

Fig. D 3 The size effect on the mean and standard deviation of creep rupture strain and rupture time with $\sigma_e = 20$ MPa

Moreover, the means and standard deviations of the creep rupture strain and time of RVEs under different stress state can be obtained as shown in Fig. D 3. Fig. D 3a and b present the mean of creep rupture strain and time obtained from MC calculation. The vertical axis of the Fig. D 3a and b includes two breaks to allow the curves to be clearly displayed within a single plot. A significant size effect or scale effect can be observed. As the RVE size increases, it tends to fail earlier. Moreover, as the stress triaxiality increases, the size effect becomes less significant. Moreover, as the RVE size increases, the standard deviation significantly decreases for different multiaxial stress states (see Fig. D 3c and d). This can be explained and predicted by the Central Limit Theorem (CLT). If the number of GBs in an RVE with edge length $a_1$ is $N_1$, with a corresponding standard deviation $\sigma_{a1}$; and the number of GBs in an RVE with edge length $a_2$ is $N_2$, with a corresponding standard deviation $\sigma_{a2}$. Then, it is obtained:

$$\sigma_{a1} \cdot \sqrt{N_1} = \sigma_{a2} \cdot \sqrt{N_2} \tag{D.1}$$

The prediction results through the CLT, shown in Fig. D 3c and d, agree well with the MC results.

2) The effect of stress state on stochastic damage

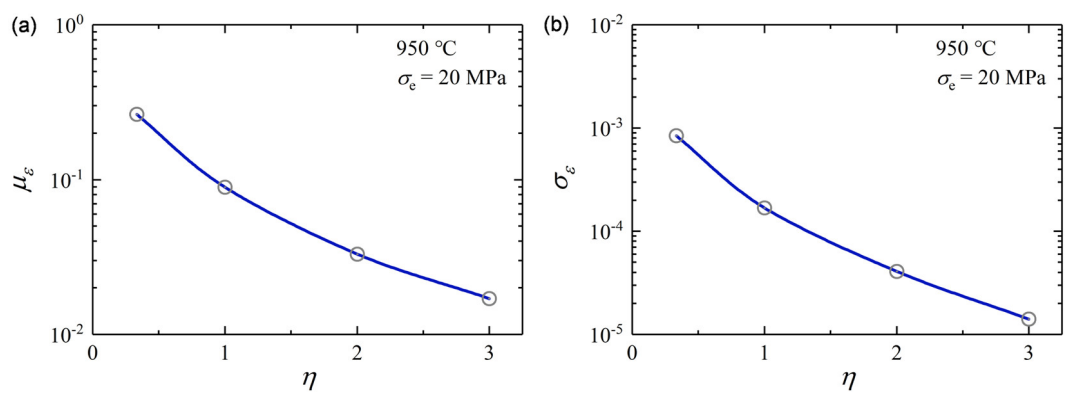

Fig. D 4 (a) The mean values of creep rupture strain versus stress triaxiality for $\sigma_e = 20$ MPa and $a = 320$ μm (b) The standard deviations of creep rupture strain versus stress triaxiality for $\sigma_e = 20$ MPa and $a = 320$ μm

The dispersion of equivalent creep rupture strain and time is also influenced by the stress state. The creep rupture strain for $\sigma_e = 20$ MPa under uniaxial and multiaxial stress states is obtained from MC approach. The creep rupture strains conform to normal distributions. The mean values and standard deviations are obtained for each case and shown in Fig. D 4a and b, respectively. Similarly, the distribution of creep rupture times for $\sigma_e = 20$ MPa under uniaxial and multiaxial stress states can also be obtained from MC approach. The corresponding mean values and standard deviations are shown in Fig. D 5a and b. It is found that the mean values of the creep rupture strain and time decreases with the increase of macro stress triaxiality. The standard deviations of creep rupture strain and time also decrease with the increase of macro stress triaxiality. The dispersion of creep rupture strain and time is caused by the differences in microstructure. The decrease of standard deviations indicates that the effect of micro structure on creep failure behavior reduces with the increasing stress triaxiality.

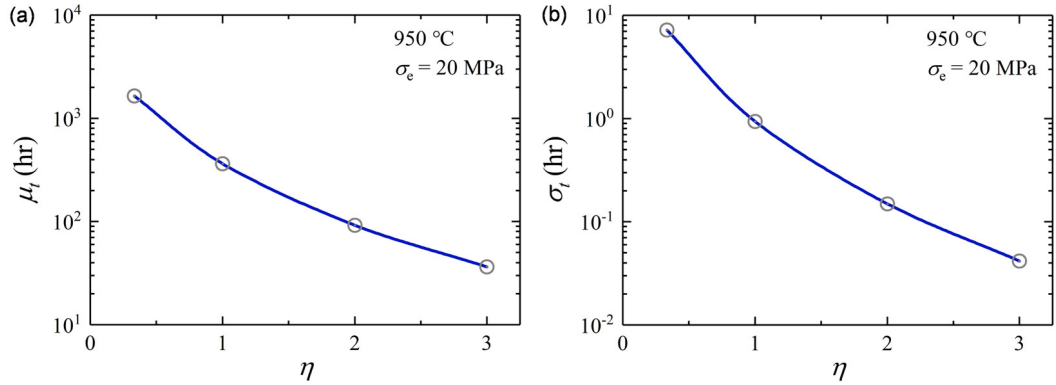

Fig. D 5 (a) The mean values of creep rupture time versus stress triaxiality for $\sigma_e = 20$ MPa and $a = 320$ μm (b) The standard deviations of creep rupture time versus stress triaxiality for $\sigma_e = 20$ MPa and $a = 320$ μm

For cases with the same equivalent stress $\sigma_e$, the increase of stress triaxiality means the increase of hydrostatic stress, leading to the increase of principal stresses. According to Eqn. (10), the increase of maximum principal stress will result in a quick creep damage rate. Hence, even though the equivalent stress $\sigma_e$ is kept same for cases shown in Fig. D 4 and Fig. D 5, the creep rupture strain and time still reduce rapidly. Due to the significant reduction in the absolute values of creep rupture strain and time, the corresponding absolute standard deviations also decrease.

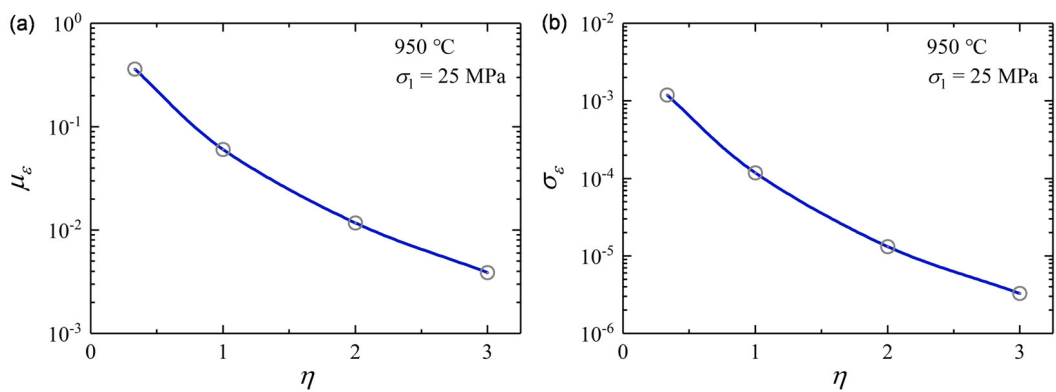

Fig. D 6 (a) The mean values of creep rupture strain versus stress triaxiality for $\sigma_1 = 25$ MPa and $a = 320$ μm (b) The standard deviations of creep rupture strain versus stress triaxiality for $\sigma_1 = 25$ MPa and $a = 320$ μm

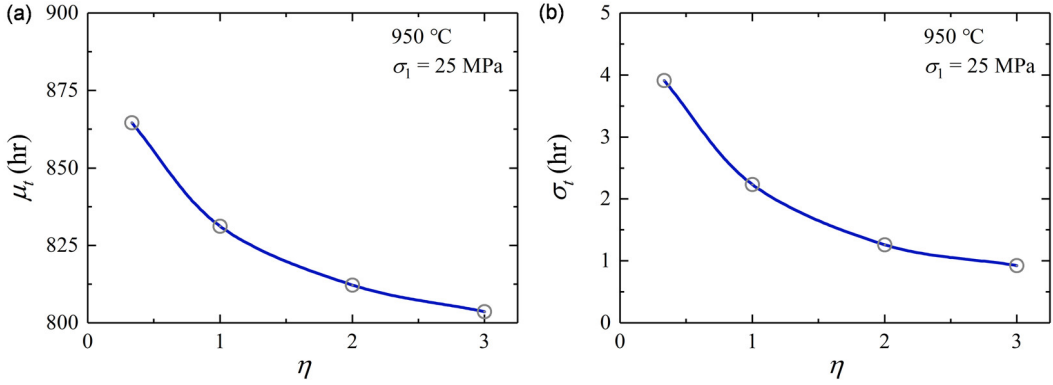

Fig. D 7 (a) The mean values of creep rupture time versus stress triaxiality $\sigma_1 = 25$ MPa and $a = 320$ μm (b) The standard deviations of creep rupture time versus stress triaxiality $\sigma_1 = 25$ MPa and $a = 320$ μm

If the maximum principal stress $\sigma_1$ keeps consistent, the creep rupture strain and time can also be obtained and presented in Fig. D 6 and Fig. D 7. It is found that the equivalent creep rupture strain significantly reduces with the increase of stress triaxiality due to the increase in volumetric strain under multiaxial stress state. However, the decrease of creep rupture time is relatively slight compared with the results shown in Fig. D 5. The reason is that the creep damage rate does not increase significantly due to the same principal stress $\sigma_1$ in Eqn. (10). The increase of stress triaxiality will result in the increase of $k$ according to Eqn. (5). According to Eqn. (10), the increase of $k$ will reduce the GB damage variation caused by the randomness of GB orientation angle $\psi_i$, leading to a reduction in the variability of $t_f$ obtained from the MC calculations. In other words, as $k$ increases, the influence of the random GB orientation angle on the variability of the damage behavior decreases. Therefore, the correlation of the orientation angle $\psi_i$ with creep rupture time reduces, resulting in a decrease in the standard deviation shown in Fig. D 7b.

# Appendix E  The FEM modelling of uniaxial tensile bar

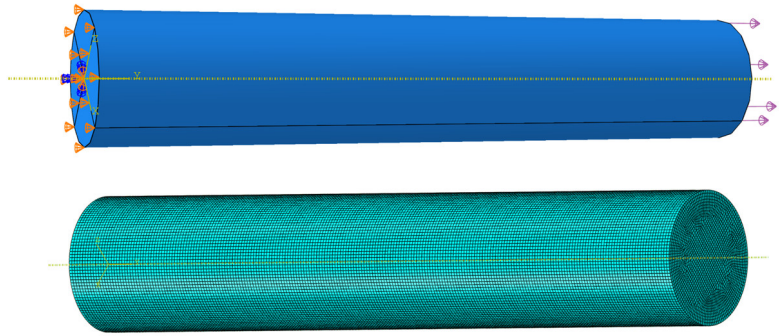

Fig. E 1 The FEM modeling and meshes of the uniaxial tensile bar

The FEM modeling and meshes of the uniaxial tensile bar are presented in Fig. E 1. The boundary conditions and the loading are presented. The radius and length are 3.175 and 34.2 mm.

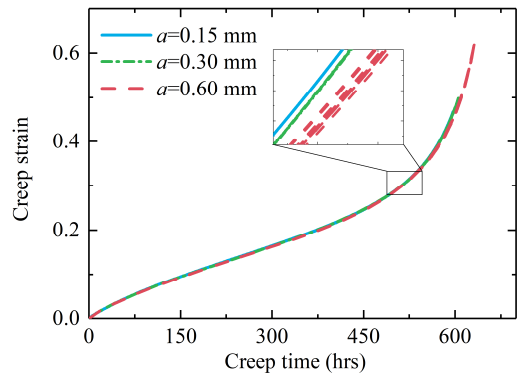

Fig. E 2 The verification of mesh convergence in uniaxial tension case

The mesh convergence in uniaxial tension case is shown in Fig. E 2. In this study, simulations were conducted with three different element sizes: 0.15 mm, 0.3 mm, and 0.6 mm. For each element size, 10 different randomly distributed cases were selected for simulation. It can be observed that the distribution of the uniaxial tensile curves is relatively small. The dispersion band of these curves gradually converges. Since the FE simulations are conducted at the structural level, the structural response can be considered as a mean behavior of the material. Therefore, the differences in uniaxial tensile creep curves with different random distributions are very minor. Overall, the smaller the mesh size is, the larger the computed creep strain will be. Due to the overall uniformity of the macroscopic stress in uniaxial tension, good mesh convergence is exhibited in the simulation.

## Appendix F The FEM modelling of pressurized tube

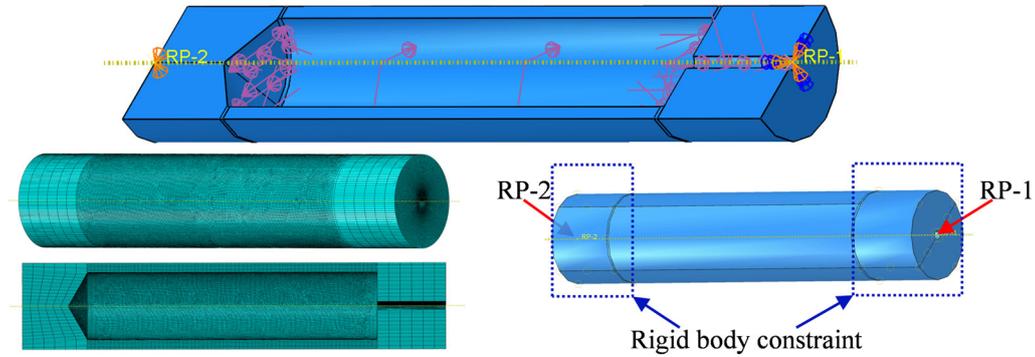

Fig. F 1 FEM modeling and meshes of the pressurized tube

As shown in Fig. F 1, the FEM modeling of the pressurized tube is conducted in the commercial software ABAQUS. Given that this study primarily focuses on the creep failure of thin-walled cylindrical tube sections, rigid constraints are applied to the ends of the tube to reduce computational costs. For the same reason, the mesh is refined along the cylindrical tube section. The element type C3D8R is employed in simulation. Displacement boundary conditions are applied to the reference points RP-1 and RP-2 corresponding to the rigid constraints. The geometric dimensions of the specimens are sourced from (Wright and wright, 2013).

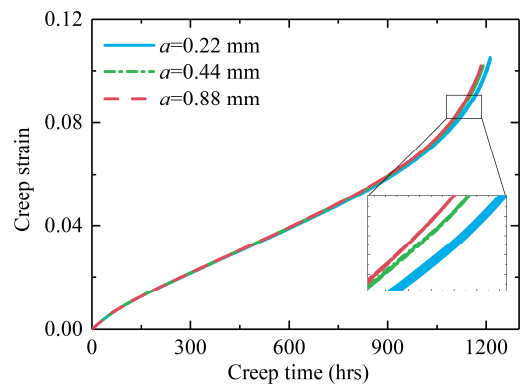

Fig. F 2 The equivalent creep strain curves calculated using elements with different sizes for pressurized tube

As shown in Fig. F 2, the mesh convergence is verified for the simulation of the pressurized tube. Similarly, simulations were conducted with three different element sizes: 0.22 mm, 0.44 mm, and 0.88 mm. For each element size, ten different randomly distributed cases were selected for simulation. Due to the presence of a significant stress gradient in pressurized tube, the curves for different element sizes exhibit noticeable differences at tertiary creep. As the mesh size decreases, both the dispersion band and the mean behavior of creep behavior tend to converge.